\PassOptionsToPackage{svgnames}{xcolor}
\documentclass[twocolumn]{aastex631}
\usepackage{CJK}

\definecolor{linkcolor}{cmyk}{1,1,0,0}

\usepackage{color,soul}
\usepackage{scalerel}
\usepackage{dashrule}
\usepackage{totcount}
\newtotcounter{citenum}
\def\oldcite{}
\let\oldcite=\bibcite
\def\bibcite{\stepcounter{citenum}\oldcite}
\usepackage{mathtools}
\usepackage[thinc]{esdiff}
\usepackage[italicdiff]{physics}

\defcitealias{RC3}{RC3}
\defcitealias{Baldwin:1981}{BPT}
\defcitealias{Sheth:2010}{S$^4$G}

\definecolor{MATLAB_blue}{rgb}{0,0.4470,0.7410}
\definecolor{MATLAB_red}{rgb}{0.8500,0.3250,0.0980}
\definecolor{MATLAB_orange}{rgb}{0.9290,0.6940,0.1250}
\definecolor{MATLAB_purple}{rgb}{0.4940,0.1840,0.5560}
\definecolor{MATLAB_green}{rgb}{0.4660,0.6740,0.1880}
\definecolor{MATLAB_cyan}{rgb}{0.3010,0.7450,0.9330}
\definecolor{MATLAB_maroon}{rgb}{0.6350,0.0780,0.1840}
\definecolor{LimeGreen}{rgb}{0.1961,0.8039,0.1961}
\definecolor{Orange}{rgb}{1,0.6471,0}

\received{November 21, 2023}
\revised{May 20, 2024}
\accepted{June 7, 2024}
\submitjournal{\textit{The Astrophysical Journal}}

\shorttitle{IMBH Candidates in Sd Galaxies}
\shortauthors{Davis et al.}

\begin{document}
\begin{CJK*}{UTF8}{gbsn}

\title{Identification of Intermediate-mass Black Hole Candidates Among a Sample of Sd Galaxies}

\correspondingauthor{Benjamin L.\ Davis}
\email{ben.davis@nyu.edu}

\author[0000-0002-4306-5950]{Benjamin L.\ Davis}
\affil{Center for Astrophysics and Space Science (CASS), New York University Abu Dhabi, PO Box 129188, Abu Dhabi, UAE}
\affil{Centre for Astrophysics and Supercomputing, Swinburne University of Technology, Hawthorn, VIC 3122, Australia}

\author[0000-0002-6496-9414]{Alister W.\ Graham}
\affil{Centre for Astrophysics and Supercomputing, Swinburne University of Technology, Hawthorn, VIC 3122, Australia}

\author[0000-0002-4622-796X]{Roberto Soria}
\affil{INAF-Osservatorio Astrofisico di Torino, Strada Osservatorio 20, I-10025 Pino Torinese, Italy}
\affil{Sydney Institute for Astronomy, School of Physics A28, The University of Sydney, Sydney, NSW 2006, Australia}
\affil{College of Astronomy and Space Sciences, University of the Chinese Academy of Sciences, Beijing 100049, China}

\author[0009-0000-2506-6645]{Zehao Jin (金泽灏)}
\affil{Center for Astrophysics and Space Science (CASS), New York University Abu Dhabi, PO Box 129188, Abu Dhabi, UAE}

\author[0000-0003-0307-4366]{Igor D.\ Karachentsev}
\affil{Special Astrophysical Observatory, Russian Academy of Sciences, N\.Arkhyz 369167, Russia}

\author{Valentina E.\ Karachentseva}
\affil{Main Astronomical Observatory of National Academy of Sciences of Ukraine, Kyiv 03143, Ukraine}

\author[0000-0003-2676-8344]{Elena D'Onghia}
\affil{Astronomy Department, University of Wisconsin, Madison, WI 53706, USA}

\begin{abstract}

We analyzed images of every northern hemisphere Sd galaxy listed in the Third Reference Catalogue of Bright Galaxies (RC3) with a relatively face-on inclination ($\theta\leq30\degr$).
Specifically, we measured the spiral arms' winding angle, $\phi$, in 85 galaxies.
We applied a novel black hole mass planar scaling relation involving the rotational velocities (from the literature) and pitch angles of each galaxy to predict central black hole masses.
This yielded 23 galaxies, each having at least a 50\% chance of hosting a central intermediate-mass black hole (IMBH), $10^2<M_\bullet\leq10^5\,\mathrm{M}_\sun$.
These 23 nearby ($\lesssim$50\,Mpc) targets may be suitable for an array of follow-up observations to check for active nuclei.
Based on our full sample of 85 Sd galaxies, we estimate that the typical Sd galaxy (which tends to be bulgeless) harbors a black hole with $\log(M_\bullet/\mathrm{M}_\sun)=6.00\pm0.14$, but with a 27.7\% chance of hosting an IMBH, making this morphological type of galaxy fertile ground for hunting elusive IMBHs.
Thus, we find that a $\sim$$10^6\,\mathrm{M}_\sun$ black hole corresponds roughly to the onset of bulge development and serves as a conspicuous waypoint along the galaxy--SMBH coevolution journey.
Our survey suggests that $>$1.22\% of bright galaxies ($B_{\rm T}\lesssim15.5$\,mag) in the local Universe host an IMBH (\textit{i.e.}, the ``occupation fraction''), which implies a number density $>$$4.96\times10^{-6}$\,Mpc$^{-3}$ for central IMBHs.
Finally, we observe that Sd galaxies exhibit an unexpected diversity of properties that resemble the general population of spiral galaxies, albeit with an enhanced signature of the eponymous prototypical traits (\textit{i.e.}, low masses, loosely wound spiral arms, and smaller rotational velocities).


\end{abstract}

\keywords{
\href{http://astrothesaurus.org/uat/1882}{Astrostatistics (1882)} ---
\href{http://astrothesaurus.org/uat/594}{Galaxy evolution (594)} ---
\href{http://astrothesaurus.org/uat/757}{Hubble classification scheme (757)} ---
\href{http://astrothesaurus.org/uat/816}{Intermediate-mass black holes (816)} ---
\href{http://astrothesaurus.org/uat/907}{Late-type galaxies (907)} ---
\href{http://astrothesaurus.org/uat/1914}{Regression (1914)} ---
\href{http://astrothesaurus.org/uat/2031}{Scaling relations (2031)} ---
\href{http://astrothesaurus.org/uat/1560}{Spiral galaxies (1560)} ---
\href{http://astrothesaurus.org/uat/1561}{Spiral pitch angle (1561)}
}

\section{Introduction}\label{sec:intro}
\end{CJK*}

Intermediate-mass black holes (IMBHs) could be characterized as the rare larval form of black holes.
Announced within a span of just three years, astronomers built upon a century of black hole research\footnote{See \citet{Graham:2016} for key results and a detailed timeline of black hole research.} to obtain emphatic evidence for the embryonic and adult forms of black holes: stellar-mass \citep{Abbott:2016} and supermassive \citep{EHT,Akiyama:2022} black holes (SMBHs), respectively.
However, speculation abounds concerning the intervening IMBH category that bridges the mass gap between stellar-mass and SMBHs.
The gap in our knowledge is akin to a boxing league that has lightweight and heavyweight divisions, but no middleweight division.
In which, there are three possibilities: (i) individual boxers have always been either lightweight or heavyweight, (ii) middleweight boxers always gain weight rapidly to become heavyweights, or (iii) there is an underground fighting league for middleweights that we are not privy to.
Thus, we are on a quest to find evidence of scenario (iii) by tracking down clandestine IMBHs in their galactic lairs.

Perhaps the first evidence of the missing demographical category of black holes came at the turn of the millennium with the identification of an IMBH with a mass $>$$700\,\mathrm{M}_\sun$ in M82 \citep{Matsushita:2000,Kaaret:2001,Ebisuzaki:2001,Matsumoto:2001}.
Still, almost a quarter-century later, obtaining direct confirmation of IMBHs is not easy \citep[see][for recent reviews]{Mezcua:2017,Koliopanos:2017b,Greene:2020}.
Even if one could prognosticate with certainty where to look for an IMBH, the telescope resources and time investment to garner definitive proof of IMBHs is significant \citep[see][their \S5]{Graham:2021}.
It is for these reasons that reconnaissance work is required to preselect the galaxies that are likely to harbor IMBHs.
Such scouting studies will become invaluable to the world's best observatories as they look for IMBHs.

Large surveys will surely play a pivotal role in the search for IMBHs.
Data mining of preexisting surveys can be a promising avenue for selection of candidates.
Furthermore, large survey telescopes like the Vera C.\ Rubin Observatory \citep[n\'{e}e Large Synoptic Survey Telescope;][]{Tyson:2002} and the Einstein Probe \citep{Yuan:2022} are expected to find vast numbers of active galactic nuclei (AGNs), tidal disruption events (TDEs), and likely IMBHs.
Careful filtering will be required to sort through the immense quantity of data produced to efficiently identify IMBH candidates.

One of the best methods for predicting black hole masses in galaxies is via black hole mass scaling relations \citep[\textit{e.g.},][]{Bennert:2011,Graham:2015,Graham:2016,D'Onofrio:2021,Izquierdo-Villalba:2023}.
By identifying galaxies that reside at the same extreme end of multiple scaling relations, the success of their combined predictive power becomes more probable \citep{Koliopanos:2017,Graham:Soria:2019,Graham:2019,Graham:2021b,Graham:2021,Davis:2020}.
Furthermore, because extrapolating is inherently uncertain, it is safer to extrapolate from multiple relations and look for agreement rather than trusting that a single relation holds beyond its defined range.
Therefore, we seek a class of galaxy that consistently occupies the extremities of separate black hole mass scaling relations.
Moreover, we effectively entwine the separate scaling relations by applying a novel trivariate relationship \citep{Davis:2023}.

Late-type spiral (\textit{e.g.}, Sd) galaxies are the most compatible morphological class of galaxy for the focus of our experiment.
Sd galaxies tick all the boxes in a census of galaxies that place them in a rare demographic category which is highly likely to cohabit with IMBHs.
We note that dwarf early-type galaxies (particularly low-mass S0 galaxies) are also good candidates to host IMBHs \citep{Graham:Soria:2019}.
Furthermore, they have been shown to occasionally possess faint disk substructure, including bars and spiral arms, the latter of which can be quantified via pitch angle measurements \citep{Jerjen:2000,Lisker:2006,Michea:2021}.
However, the oft-hidden volute structure (\emph{if} it exists) requires significant image processing to extract the embedded disk component, which itself has been historically missed even in bright early-type galaxies.
Specifically, \citet{Lisker:2006} found that 41 out of the 476 ($8.6\%$) dwarf early-type galaxies in their sample showed ``possible, probable, or unambiguous disk features.''
Although, the semblance of spiral structure in dwarf early-type galaxies could be caused by tidal triggering resulting from the cluster harassment of passive dwarf galaxies \citep{SmithR:2021}.

Sd galaxies have open spiral structures, low total masses, small central velocity dispersions, slow rotational velocities, and are likely bulgeless; all such traits establish environments that should be ripe for the existence of IMBHs.
Moreover, their structure is regular enough to not place them in peculiar or Magellanic type distinctions.
The preponderance of low-mass, disk-dominated, bulgeless galaxies in our sample also distinguish our galaxies as having lived relatively merger-free lives.
This creates an ideal scenario to conduct a clean test of IMBH growth in the absence of external influences on nuclear black holes that must have evolved in relative isolation (unlike Sm galaxies).
Indeed, it is expected that the galaxy merger fraction monotonically increases as a function of stellar mass \citep{Guzman:2023} and bulge mass \citep{Graham:2023}.
Nonetheless, evidence shows that non-merger processes alone are sufficient to fuel massive black hole growth in galaxies \citep{Smethurst:2021} and sustain coevolution between SMBHs and their host galaxies \citep{Smethurst:2022}.

The ongoing pursuit of IMBHs has seen significant recent contributions.
Using the \textit{Chandra} X-ray Observatory \citep[CXO;][]{Weisskopf:2000}, the ``Chandra Virgo Cluster Survey of Spiral Galaxies'' \citep[][see also \citealt{Chilingarian:2018} and \citealt{Bi:2020}]{Soria:2022} obtained long exposures for spiral galaxies in the Virgo Cluster.
When combined with archival data, this project has X-ray imaging for all spiral galaxies in the Virgo Cluster with star-formation rates $\gtrsim$$1\,\mathrm{M}_\sun$\,yr$^{-1}$.
Early identification of nuclear X-ray point sources from the archival CXO data, coupled with black hole mass scaling relations, yielded $3+11$ strong\footnote{Here, \emph{strong} candidates are those with both a predicted black hole mass $\lesssim$10$^5$\,M$_\sun$ and a centrally-located X-ray point source.} IMBH candidates \citep{Graham:2019,Graham:2021}.
\citet{Karachentsev:2019} collated a catalog of 220 face-on bulgeless galaxies, approximately half of which exhibit unresolved nuclei.
Because nuclear star clusters (NSCs) are known to scale with their host galaxies \citep{Balcells:2003,Graham:2003,Wehner:2006,Brok:2014,Georgiev:2016,Janssen:2019,Pechetti:2020} and their central black holes \citep{Graham:2009,Scott:2013,Georgiev:2016,Graham:2016b,Graham:2019c,Neumayer:2020}, such a catalog is a valuable reference of potential IMBH host galaxies.

Sd galaxies are certainly not the only place where IMBHs might exist.
\citet{McKernan:2012} envision an efficient process by which IMBHs may be efficiently grown in AGNs in the disks surrounding SMBHs.
There might be IMBHs roaming our Galaxy \citep{Schodel:2005,Oka:2017,Tsuboi:2017,Tsuboi:2019,Tsuboi:2020,Takekawa:2019,Takekawa:2020,Zhu:2020,Reid:2020,GRAVITY:2020,Weller:2022} or in its satellites that reveal themselves more readily than IMBHs in extragalactic environments.\footnote{
Specifically, \citet{Paynter:2021} estimated there are $\approx$$4.6\times10^4$ IMBHs with masses between $\approx$$10^4$--$10^5\,\mathrm{M_\sun}$ in the neighborhood of the Milky Way.}
Indeed, \citet{Nguyen:2019} estimated a black hole mass ($M_\bullet\equiv M_\mathrm{BH}$) of $\log(M_\bullet/\mathrm{M}_\sun)=3.83_{-0.60}^{+0.43}$ in \object{Messier~110} (a dwarf elliptical satellite of the Andromeda Galaxy) via stellar dynamical modeling.
Verily, not all fish are in the oceans but also rivers, ponds, and lakes.
Analogously, Sd galaxies represent the major fishing areas for catching potential IMBHs.
It is our endeavor here to establish: in general, (i) how common IMBHs might be in Sd galaxies and explicitly, (ii) which Sd galaxies pose the most promising potential for future studies to search for IMBHs.

We define our sample of Sd galaxies in \S\ref{sec:sample}, and apply a planar black hole mass scaling relation to the collection of galaxies (in \S\ref{sec:masses}).
In \S\ref{sec:results}, we present the results and statistics for our sample and identify our targets of interest.
Additionally, we discuss other observational considerations such as AGNs and X-ray point sources.
Finally, we provide a discussion (in \S\ref{sec:discussion}) on our findings for Sd galaxies, detail the implications from IMBH research, consider the prospects of detecting IMBHs in our sample, consider the outlook for follow-up investigations, summarize our findings in \S\ref{sec:conclusions}.
We represent masses ($M$) throughout this work as logarithmic (solar) masses ($\mathcal{M}$), such that $\mathcal{M}\equiv\log(M/\mathrm{M}_\sun)$.
Stellar masses have all been adjusted to conform with the \citet{Chabrier:2003} initial mass function\footnote{
See \citet{Davis:2018,Davis:2019b} for further details on how we homogenize stellar masses from various studies.}
and we assume cosmographic parameters from \citet{Planck:2020}.
All uncertainties are quoted at $1\,\sigma \equiv 68.3\%$ confidence intervals; median absolute deviations are given as uncertainties associated with medians.

\section{Sample of Galaxies}\label{sec:sample}

We assembled a sample of Sd (\textit{i.e.}, SAd, SABd, and SBd) type galaxies \citep{Shapley:1940,Vaucouleurs:1959,Graham:2019b}.
We focussed on the Sd, rather than Sc and earlier types, because of the expected larger spiral-arm winding angles; open spiral arms correlate with low black hole masses \citep{Seigar:2008,Berrier:2013,Davis:2017}.
The Magellanic-like Sm, and Sdm, (nor irregular) galaxies were not included because of their inherent disrupted and asymmetric structures.\footnote{
Nevertheless, there is evidence that IMBHs may reside in irregular and/or Magellanic-type morphologies \citep[\textit{e.g.}, NGC~5408;][]{Strohmayer:2009}.
}
Sd galaxies exhibit open spiral-arm structures and possess very faint (or no) bulges; they are ideal galaxies to host IMBHs.
Because the bulge-to-total flux ratio ($B/T$) of Sd galaxies are low enough\footnote{Observed in the $B$-band, the bulge-to-total flux ratio for Sd galaxies has been reported as $0.029$ \citep{Simien:1986} and $0.027_{-0.016}^{+0.066}$ \citep{Graham:2008}.} to be considered ``bulgeless,''\footnote{For Sd galaxies, which typically lack conventional bulges, the central component proxy is likely a weak pseudobulge or ``barge'' (portmanteau of ``bar'' and ``bulge'').} traditional (black hole mass)--(bulge mass) and (black hole mass)--(S\'ersic index) relations are problematic, indicating a population of galaxies that likely possess small (or no) nuclear black holes.\footnote{
See \citet{Bohn:2020} for a thorough list and discussion of known bulgeless galaxies with black hole mass estimates.}

Bulgeless galaxies may still be analyzed with other black hole mass scaling relations that are not dependent on bulge masses, but some other physical property, such as spiral-arm pitch angle \citep{Seigar:2008,Berrier:2013,Davis:Thesis,Davis:2017}, central velocity dispersion \citep{Ferrarese:2000,Gebhardt:2000,Sahu:2019}, total stellar mass \citep{Beifiori:2012,Davis:2018,Sahu:2019b,Graham:2023},\footnote{Moreover, the fact that black holes follow the (black hole mass)--(host stellar mass) relation even in bulgeless galaxies indicates that massive black holes may form before stellar bulges in galaxies \citep[\textit{e.g.},][]{Chen:2023}.} and rotational velocity \citep{Ferrarese:2002,Sabra:2015,Davis:2019,Smith:2021}.
The total number of globular clusters bound to a galaxy is also an intriguing proxy for its central black hole mass \citep{Burkert:2010,Harris:2011,Harris:2014,Gonzalez:2022}.
Notably, \citet{Bluck:2023} find that the stellar gravitational potential of a galaxy is strongly correlated with SMBH mass, which is easy to measure in both observational and simulated data.
The measurement of spiral-arm pitch angle ($\phi$) requires a spiral galaxy first to be corrected to a face-on orientation via a deprojection of its apparent inclination with respect to the plane of the sky \citep[see][figure~2]{Davis:2012}.
Because of this necessity, face-on galaxies are far easier to measure, with pitch angles becoming progressively harder to measure for more inclined galaxies and impossible for edge-on galaxies.
For this reason, we have selected our sample of spiral galaxies to be close to face-on.
Additionally, dust/inclination effects are minimized for galaxies when they are near to face-on.

We have constructed this study to take advantage of the remarkably low level of scatter for black hole mass estimates from the $M_\bullet$--$\phi$ relation.
The usual go-to black hole mass scaling relation for the majority of studies has been the $M_\bullet$--$\sigma_0$ relation, due to its low level of scatter and availability of central velocity dispersion ($\sigma_0$) measurements.
Indeed, central velocity dispersion is often described as the most fundamental black hole mass scaling relation parameter \citep[\textit{e.g.},][]{Nicola:2019}.
However, this is predominantly the case for early-type galaxies; for late-type galaxies, the scatter is higher and $\sigma_0$ measurements are less common.
From a common sample of 44 spiral galaxies, \citet{Davis:2017} showed that the intrinsic scatter for the $M_\bullet$--$\sigma_0$ relation is 0.21\,dex higher than that of the $M_\bullet$--$\phi$ relation.

For the selection of our galaxies, we turned to the Third Reference Catalogue of Bright Galaxies \citep[][hereafter \citetalias{RC3}]{RC3}.
The full \citetalias{RC3} sample contains 23,022 galaxies (17,801 with morphological classifications) with apparent diameters larger than one arcminute ($D_{25}>1\arcmin$), total $B$-band magnitudes brighter than about 15.5\,mag (Vega $B_{\rm T}\lesssim15.5$\,mag), and recessional velocities $cz<1$5,000\,km\,s$^{-1}$ ($z<0.050$; luminosity distances out to $d=230$\,Mpc).
Using their online database,\footnote{\url{https://heasarc.gsfc.nasa.gov/W3Browse/all/rc3.html}} we selected a sample consisting of all Sd galaxies (Hubble sequence morphological stage, $T=7.0$)\footnote{In their study of spiral galaxies in the \citetalias{RC3}, \citet{Ma:1999} indeed found that Sd types had the highest observed pitch angles, $|\bar{\phi}|=25\fdg00$.} with inclination angles ($\theta$) such that $0\degr\leq \theta \leq 30\degr$, and in the northern celestial hemisphere (declinations greater than $0\degr$).
Here, the inclination angle is determined from the mean ratio ($\log{\rm R_{25}}$) of the major isophotal diameter (${\rm D_{25}}$) to the minor isophotal diameter (${\rm d_{25}}$) measured at or reduced to the $B$-band surface brightness level $\mu_B=25.0{\rm \,mag\,arcsec^{-2}}$.
Therefore, we selected $0\leq\log{\rm R_{25}}\lesssim0.06$ because $\theta\equiv \sec^{-1}({\rm R_{25}})$.

These selection criteria yielded a sample of 85 spiral galaxies (see Table~\ref{tab:sample}).
We chose to restrict our sample to the northern sky so that our entire sample would be visible by extensive sky surveys such as Pan-STARRS1 and SDSS to assure uniform access to high-quality imaging of volute structure to facilitate the measurement of spiral-arm pitch angles.
Therefore, our sample should represent roughly $1/15$ of the entire \citetalias{RC3} sample of Sd galaxies (\textit{i.e.}, half of the sky and $1-\cos30\degr$ of all random inclination orientations).
In actuality, there are 787 Sd galaxies in the \citetalias{RC3}, thus we have selected $85/787\approx11\%$, \textit{cf.}\ $1/15\approx7\%$.

\startlongtable
\begin{deluxetable*}{lrccrrrrllrrr}
\tabletypesize{\tiny}
\tablecolumns{13}
\tablecaption{Sample of 85 Sd Galaxies}\label{tab:sample}
\tablehead{
\colhead{} & \colhead{} & \colhead{} & \multicolumn{2}{c}{CXO} & \colhead{} & \colhead{} & \colhead{} & \colhead{} & \colhead{} & \colhead{} & \colhead{} & \colhead{} \\
\cline{4-5}
\colhead{Galaxy} & \colhead{Distance} & \colhead{\citetalias{Baldwin:1981}} & \colhead{Exp} & \colhead{\#} & \colhead{$|\phi|$} & \colhead{$\sigma_0$} & \colhead{$\mathcal{M}_{\rm gal,\star}$} & \colhead{$\theta$} & \colhead{$v_\mathrm{max}$} & \colhead{$\mathcal{M}_\bullet$} & \colhead{$P(\mathcal{M}_\bullet\leq5)$} & \colhead{$n\,\sigma$} \\
\colhead{} & \colhead{[Mpc]} & \colhead{} & \colhead{[ks]} & \colhead{} & \colhead{[$\degr$]} & \colhead{$\left[\frac{\text{km}}{{\text{s}}}\right]$} & \colhead{[dex]} & \colhead{[$\degr$]} & \colhead{$\left[\frac{\text{km}}{{\text{s}}}\right]$} & \colhead{[dex]} & \colhead{[\%]} & \colhead{[$\sigma$]} \\
\colhead{(1)} & \colhead{(2)} & \colhead{(3)} & \colhead{(4)} & \colhead{(5)} & \colhead{(6)} & \colhead{(7)} & \colhead{(8)} & \colhead{(9)} & \colhead{(10)} & \colhead{(11)} & \colhead{(12)} & \colhead{(13)}
}
\startdata
\object{IC~951} & $62.2\pm7.2$\phn	& \ion{H}{2} & \nodata & \nodata & $9.5\pm2.9$ & \nodata & $10.2\pm0.2$ & $17\pm2$ ($0\pm44$) & $199\pm29$ & $7.6\pm0.4$ & $\approx$0 & $-5.9$ \\
\object{IC~1221} & $74.9\pm6.5$\phn & \ion{H}{2} & 5 & 0 & $22.7\pm3.9$ & \nodata & $10.3\pm0.1$ & $18\pm1$ ($27\pm13$) & \phn$62\pm5$ & $4.2\pm0.5$ & 93.3 & 1.5 \\
\object{IC~1774} & $41.8\pm5.8$\phn & \nodata & \nodata & \nodata & $22.8\pm2.4$ & \nodata & $9.9\pm0.2$ & $43\pm2$ ($21\pm17$) & \phn$99\pm5$ & $5.0\pm0.4$ & 49.7 & 0.0 \\
\object{IC~1776} & $40.9\pm6.1$\phn & \nodata & \nodata & \nodata & $17.9\pm3.0$ & \nodata & $9.9\pm0.2$ & $28\pm1$ ($12\pm30$) & \phn$99\pm3$ & $5.6\pm0.4$ & 8.0 & $-1.4$ \\
\object{NGC~2500} & $12.0\pm9.4$\phn & \nodata & 3 & 7 & $16.3\pm1.3$ & \nodata & $9.6\pm0.7$ & $28\pm11$ ($24\pm9$) & \phn$90\pm34$ & $5.6\pm0.7$ & 20.7 & $-0.8$ \\
\object{NGC~2657} & $59.0\pm7.2$\phn & \nodata & \nodata & \nodata & $15.8\pm4.5$ & \nodata & $10.4\pm0.2$ & $35\pm3$ ($12\pm36$) & $144\pm11$ & $6.4\pm0.5$ & 0.4 & $-2.6$ \\
\object{NGC~3906} & $19.0\pm8.0$\phn & \nodata & 5 & 0 & $16.6\pm1.9$ & \nodata & $9.6\pm0.4$ & $42\pm4$ ($27\pm10$) & $134\pm12$ & $6.2\pm0.3$ & $\approx$0 & $-3.6$ \\
\object{NGC~3913} & $17.9\pm7.5$\phn & \nodata & 5 & 0 & $12.9\pm0.2$ & \nodata & $9.5\pm0.4$ & $16\pm4$ ($12\pm18$) & \phn$69\pm19$ & $5.5\pm0.5$ & 19.0 & $-0.9$ \\
\object{NGC~4393} & $7.7\pm1.3$\phn & \nodata & 5 & 0 & $9.7\pm2.8$ & \nodata & $8.6\pm0.2$ & $56\pm1$ ($21\pm17$) & \phn$62\pm2$ & $5.6\pm0.4$ & 4.7 & $-1.7$ \\
\object{NGC~5148} & $87.8\pm6.5$\phn & \nodata & 7 & 0 & $18.7\pm3.7$ & \nodata & $10.1\pm0.1$ & $25\pm1$ ($21\pm17$) & \phn$84\pm6$ & $5.2\pm0.5$ & 34.5 & $-0.4$ \\
\object{NGC~5668} & $21.3\pm6.2$\phn & \nodata & \nodata & \nodata & $27.3\pm2.8$ & $53\pm8$ & $10.1\pm0.3$ & $15\pm1$ ($24\pm9$) & $152\pm8$ & $5.2\pm0.4$ & 31.2 & $-0.5$ \\
\object{NGC~6617} & $91.4\pm6.5$\phn & \nodata & \nodata & \nodata & $13.3\pm1.9$ & \nodata & $10.2\pm0.2$ & $52\pm1$ ($29\pm12$) & $144\pm6$ & $6.7\pm0.3$ & $\approx$0 & $-5.5$ \\
\object{NGC~6687} & $44.3\pm6.0$\phn & \nodata & \nodata & \nodata & $24.5\pm0.6$ & \nodata & $9.8\pm0.2$ & $42\pm0$ ($12\pm36$) & $101\pm19$\tablenotemark{$^\dagger$} & $4.9\pm0.4$ & 63.5 & 0.3 \\
\object{NGC~7363} & $91.2\pm6.6$\phn & \nodata & \nodata & \nodata & $10.8\pm1.3$ & \nodata & $10.5\pm0.2$ & $29\pm1$ ($21\pm21$) & $242\pm14$ & $7.8\pm0.3$ & $\approx$0 & $-10.1$ \\
\object{NGC~7437} & $26.0\pm5.9$\phn & \nodata & \nodata & \nodata & $18.0\pm1.4$ & \nodata & $9.6\pm0.3$ & $17\pm1$ ($0\pm44$) & $153\pm13$ & $6.3\pm0.3$ & $\approx$0 & $-4.2$ \\
\object{NGC~7535} & $58.5\pm6.2$\phn & \nodata & \nodata & \nodata & $13.2\pm1.0$ & \nodata & $10.2\pm0.1$ & $28\pm2$ ($0\pm44$) & $109\pm8$\phn & $6.2\pm0.3$ & $\approx$0 & $-4.3$ \\
\object{UGC~42} & $69.2\pm6.3$\phn & \nodata & \nodata & \nodata & $16.0\pm4.8$ & \nodata & $9.5\pm0.2$ & $41\pm6$ ($0\pm104$) & \phn$95\pm11$ & $5.7\pm0.6$ & 12.2 & $-1.2$ \\
\object{UGC~283} & $47.5\pm6.0$\phn & \nodata & \nodata & \nodata & $12.6\pm4.0$ & \nodata & $9.8\pm0.2$ & $45\pm1$ ($21\pm17$) & $144\pm9$\phn & $6.8\pm0.5$ & $\approx$0 & $-3.6$ \\
\object{UGC~336} & $75.0\pm6.5$\phn & \nodata & \nodata & \nodata & $16.4\pm3.1$ & \nodata & $10.0\pm0.2$ & $35\pm3$ ($21\pm14$) & $180\pm16$ & $6.7\pm0.4$ & $\approx$0 & $-4.1$ \\
\object{UGC~384} & $50.9\pm5.3$\phn & \nodata & \nodata & \nodata & $19.0\pm0.7$ & \nodata & $10.0\pm0.2$ & $45\pm2$ ($21\pm17$) & $121\pm6$\phn & $5.8\pm0.3$ & 0.1 & $-3.1$ \\
\object{UGC~1341} & $142.2\pm6.6$\phn & \nodata & \nodata & \nodata 	& $10.8\pm2.8$ & \nodata & $10.5\pm0.1$ & $32\pm3$ ($21\pm17$) & $208\pm18$ & $7.6\pm0.4$ & $\approx$0 & $-6.6$ \\
\object{UGC~1544} & $55.5\pm5.4$\phn & \nodata & \nodata & \nodata 	& $33.1\pm1.9$ & \nodata & $9.6\pm0.2$	& $53\pm7$ ($29\pm14$) & \phn$49\pm5$\phn & $2.5\pm0.4$ & $\approx$100 & 6.3 \\
\object{UGC~1606} & $104.9\pm6.6$\phn & \nodata & \nodata & \nodata 	& $11.4\pm4.7$ & \nodata & $9.9\pm0.2$ & $39\pm2$ ($24\pm18$) & $152\pm8$\phn & $7.0\pm0.5$ & $\approx$0 & $-3.7$ \\
\object{UGC~1702} & $163.5\pm6.6$\phn & \nodata & \nodata & \nodata 	& $5.8\pm2.8$ & \nodata & $10.6\pm0.2$ & $20\pm1$ ($21\pm17$) & $284\pm14$ & $8.6\pm0.4$ & $\approx$0 & $-9.8$ \\
\object{UGC~1795} & $187.1\pm6.6$\phn & \nodata & \nodata & \nodata 	& $23.3\pm3.6$ & \nodata & $10.6\pm0.2$ & $59\pm3$ ($27\pm31$) & $226\pm18$ & $6.4\pm0.5$ & 0.3 & $-2.8$ \\
\object{UGC~1833} & $52.1\pm5.4$\phn & \nodata & \nodata & \nodata & $25.3\pm2.1$ & \nodata & $9.9\pm0.2$ & $61\pm3$ ($21\pm21$) & \phn$82\pm4$\phn & $4.4\pm0.3$ & 95.6 & 1.7 \\
\object{UGC~1897} & $109.0\pm6.6$\phn & \nodata & \nodata & \nodata 	& $28.7\pm4.9$ & \nodata & $10.2\pm0.2$ & $40\pm1$ ($27\pm13$) & $132\pm6$\phn & $4.8\pm0.7$ & 61.8 & 0.3 \\
\object{UGC~2008} & $55.8\pm5.5$\phn & \nodata & \nodata & \nodata & $21.6\pm3.5$ & \nodata & $9.4\pm0.2$ & $28\pm3$ ($0\pm61$) & $109\pm12$ & $5.3\pm0.5$ & 26.3 & $-0.6$ \\
\object{UGC~2109} & $39.6\pm5.3$\phn & \nodata & \nodata & \nodata & $17.0\pm2.0$ & \nodata & $10.0\pm0.2$ & $39\pm2$ ($29\pm9$) & $112\pm5$\phn & $5.9\pm0.3$ & 0.4 & $-2.7$ \\
\object{UGC~2437} & $40.1\pm5.9$\phn & \nodata & \nodata & \nodata & $31.8\pm3.7$ & \nodata & $10.0\pm0.2$ & $39\pm2$ ($24\pm18$) & $131\pm9$\phn & $4.4\pm0.6$ & 86.4 & 1.1 \\
\object{UGC~2458} & $218.3\pm6.6$\phn & \nodata & \nodata & \nodata & $11.9\pm0.9$ & \nodata & $10.7\pm0.2$ & $36\pm2$ ($24\pm35$) & $175\pm34$\tablenotemark{$^\dagger$} & $7.2\pm0.4$ & $\approx$0 & $-5.2$ \\
\object{UGC~2623} & $47.5\pm5.3$\phn & \nodata & \nodata & \nodata & $11.6\pm3.0$ & \nodata & $9.8\pm0.2$ & $35\pm4$ ($24\pm18$) & \phn$81\pm9$\phn & $5.9\pm0.4$ & 2.1 & $-2.0$ \\
\object{UGC~2671} & $90.6\pm6.4$\phn & \nodata & \nodata & \nodata & $16.6\pm1.7$ & \nodata & $10.1\pm0.2$ & $25\pm4$ ($0\pm104$) & $187\pm31$ & $6.8\pm0.4$ & $\approx$0 & $-4.4$ \\
\object{UGC~2935} & $56.8\pm6.2$\phn & \nodata & \nodata & \nodata & $24.1\pm2.9$ & \nodata & $9.1\pm0.2$ & $62\pm2$ ($0\pm53$) & \phn$33\pm2$\phn & $3.0\pm0.4$ & $\approx$100 & 4.9 \\
\object{UGC~3074} & $52.7\pm6.1$\phn & \nodata & \nodata & \nodata & $17.3\pm0.2$ & \nodata & $9.9\pm0.2$ & $23\pm4$ ($27\pm13$) & $121\pm22$ & $6.0\pm0.4$ & 0.6 & $-2.5$ \\
\object{UGC~3364} & $59.2\pm6.4$\phn & \nodata & \nodata & \nodata & $16.5\pm2.7$ & \nodata & $9.7\pm0.2$ & $33\pm3$ ($0\pm61$) & $109\pm9$\phn & $5.9\pm0.4$ & 1.3 & $-2.2$ \\
\object{UGC~3402} & $222.9\pm6.6$\phn & \nodata & \nodata & \nodata & $16.6\pm1.3$ & \nodata	& $10.9\pm0.2$ & $34\pm1$ ($27\pm13$) & $202\pm40$\tablenotemark{$^\dagger$} & $6.9\pm0.4$ & $\approx$0 & $-4.5$ \\
\object{UGC~3702} & $65.8\pm6.6$\phn & \nodata & \nodata & \nodata & $16.2\pm2.8$ & \nodata & $10.0\pm0.2$ & $27\pm1$ ($0\pm53$) & \phn$65\pm4$\phn & $5.0\pm0.4$ & 48.0 & 0.0 \\
\object{UGC~3799} & $81.5\pm6.7$\phn & \nodata & \nodata & \nodata & $10.7\pm1.4$ & \nodata & $10.4\pm0.2$ & $39\pm1$ ($27\pm13$) & $144\pm5$\phn & $6.9\pm0.3$ & $\approx$0 & $-7.0$ \\
\object{UGC~3826} & $26.8\pm7.0$\phn & \nodata & \nodata & \nodata & $19.9\pm4.8$ & \nodata & $9.3\pm0.3$ & $40\pm4$ ($29\pm9$) & \phn$32\pm3$\phn & $3.4\pm0.6$ & 99.6 & 2.7 \\
\object{UGC~3875} & $71.6\pm6.7$\phn & \nodata & \nodata & \nodata & $21.4\pm3.8$ & \nodata & $10.0\pm0.2$ & $30\pm5$ ($29\pm12$) & $171\pm25$ & $6.1\pm0.5$ & 2.0 & $-2.0$ \\
\object{UGC~3949} & $89.9\pm6.9$\phn & \nodata & \nodata & \nodata & $25.8\pm0.3$ & $70\pm5$\tablenotemark{$^\ast$} & $10.5\pm0.1$ & $32\pm1$ ($0\pm44$) & $101\pm6$\phn & $4.7\pm0.3$ & 87.6 & 1.2 \\
\object{UGC~4077} & $62.0\pm6.7$\phn & \ion{H}{2} & \nodata & \nodata & $31.1\pm1.1$ & \nodata & $10.0\pm0.2$ & $33\pm3$ ($17\pm25$) & $130\pm12$ & $4.5\pm0.3$ & 96.0 & 1.7 \\
\object{UGC~4363} & $48.8\pm6.5$\phn & \nodata & \nodata & \nodata & $17.6\pm4.2$ & \nodata & $9.8\pm0.1$ & $76\pm1$ ($12\pm24$) & \phn$36\pm2$\phn & $3.8\pm0.5$ & 98.7 & 2.2 \\
\object{UGC~4622} & $175.9\pm6.8$\phn & Comp & \nodata & \nodata & $23.4\pm2.3$ & $107\pm8$\tablenotemark{$^\ast$} & $10.9\pm0.1$ & $19\pm1$ ($21\pm21$) & $201\pm34$\tablenotemark{$^\dagger$} & $6.2\pm0.5$ & 0.6 & $-2.5$ \\
\object{UGC~4831} & $62.2\pm7.0$\phn & \ion{H}{2} & \nodata & \nodata & $15.9\pm0.7$ & \nodata & $9.8\pm0.2$ & $26\pm3$ ($0\pm44$) & $127\pm16$ & $6.2\pm0.3$ & $\approx$0 & $-3.7$ \\
\object{UGC~5142} & $98.1\pm7.3$\phn & \nodata & \nodata & \nodata & $9.9\pm2.6$ & \nodata & $10.1\pm0.2$ & $36\pm3$ ($0\pm104$) & $103\pm8$\phn & $6.5\pm0.4$ & $\approx$0 & $-3.9$ \\
\object{UGC~5344} & $62.0\pm7.5$\phn & \ion{H}{2} & \nodata & \nodata & $15.3\pm3.4$ & \nodata & $9.9\pm0.2$ & $28\pm5$ ($17\pm21$) & \phn$37\pm6$\phn & $4.1\pm0.5$ & 95.7 & 1.7 \\
\object{UGC~5460} & $20.6\pm8.0$\phn & \nodata & 15 & 0 & $26.2\pm2.2$ & \nodata & $9.4\pm0.4$ & $54\pm4$ ($17\pm21$) & \phn$46\pm3$\phn & $3.3\pm0.4$ & $\approx$100 & 4.6 \\
\object{UGC~6505} & $97.6\pm6.8$\phn & \nodata & \nodata & \nodata & $10.0\pm4.7$ & \nodata & $10.0\pm0.2$ & $19\pm4$ ($17\pm21$) & $115\pm21$\tablenotemark{$^\dagger$} & $6.6\pm0.6$ & 0.4 & $-2.7$ \\
\object{UGC~6616} & $20.5\pm7.4$\phn & \nodata & \nodata & \nodata & $28.6\pm4.3$ & \nodata & $9.3\pm0.4$ & $30\pm1$ ($17\pm25$) & \phn$90\pm5$\phn & $4.2\pm0.6$ & 92.3 & 1.4 \\
\object{UGC~6893} & $84.2\pm6.6$\phn & \nodata & \nodata & \nodata & $15.9\pm4.1$ & \nodata & $10.0\pm0.1$ & $17\pm8$ ($24\pm15$) & $187\pm80$ & $6.9\pm0.9$ & 1.7 & $-2.1$ \\
\object{UGC~7942} & $7.0\pm3.7$\phn & \nodata & \nodata & \nodata & $16.5\pm1.3$ & \nodata & $8.1\pm0.5$ & $31\pm14$ ($21\pm10$) & \phn$35\pm12$\tablenotemark{$^\dagger$} & $3.9\pm0.6$ & 95.7 & 1.7 \\
\object{UGC~8153} & $42.5\pm7.2$\phn & \ion{H}{2} & \nodata & \nodata & $31.6\pm3.5$ & \nodata & $9.8\pm0.2$ & $29\pm3$ ($27\pm13$) & $111\pm9$\phn & $4.1\pm0.5$ & 94.4 & 1.6 \\
\object{UGC~8171} & $292.5\pm7.4$\phn & AGN & \nodata & \nodata & $22.7\pm1.3$ & \nodata & $11.2\pm0.1$ & $25\pm4$ ($21\pm17$) & $240\pm40$\tablenotemark{$^\dagger$} & $6.6\pm0.4$ & $\approx$0 & $-3.9$ \\
\object{UGC~8436} & $45.8\pm7.3$\phn & \ion{H}{2} & \nodata & \nodata & $10.6\pm1.2$ & \nodata & $9.6\pm0.2$ & $24\pm8$ ($21\pm17$) & \phn$71\pm23$ & $5.7\pm0.6$ & 11.4 & $-1.2$ \\
\object{UGC~8611} & $41.4\pm7.1$\phn & \ion{H}{2} & \nodata & \nodata 	& $16.4\pm4.3$ & \nodata & $9.5\pm0.2$ & $43\pm4$ ($27\pm13$) & $102\pm8$\phn & $5.8\pm0.5$ & 7.6 & $-1.4$ \\
\object{UGC~8637} & $90.7\pm6.4$\phn & \ion{H}{2} & \nodata & \nodata & $24.7\pm3.3$ & \nodata & $10.3\pm0.1$ & $29\pm7$ ($21\pm17$) & \phn$97\pm21$ & $4.8\pm0.6$ & 65.5 & 0.4 \\
\object{UGC~8670} & $271.0\pm7.4$\phn & \nodata 	& \nodata 	& \nodata 	& $20.9\pm3.3$ & \nodata & $11.1\pm0.1$ & $18\pm7$ ($0\pm44$) & $238\pm39$\tablenotemark{$^\dagger$} & $6.7\pm0.5$ & $\approx$0 & $-3.4$ \\
\object{UGC~9008} & $72.3\pm6.6$\phn & \ion{H}{2} & \nodata & \nodata & $11.5\pm4.6$ & \nodata & $9.9\pm0.1$ & $25\pm5$ ($0\pm53$) & $130\pm27$ & $6.7\pm0.6$ & 0.4 & $-2.7$ \\
\object{UGC~9010} & $98.6\pm6.5$\phn & \nodata & \nodata & \nodata & $13.6\pm0.9$ & \nodata & $10.1\pm0.2$ & $46\pm3$ ($29\pm12$) & $150\pm11$ & $6.7\pm0.3$ & $\approx$0 & $-6.3$ \\
\object{UGC~9042} & $108.9\pm6.5$\phn & \ion{H}{2} & \nodata & \nodata & $11.0\pm2.8$ & \nodata & $9.9\pm0.2$ & $20\pm5$ ($29\pm12$) & $229\pm59$ & $7.7\pm0.6$ & $\approx$0 & $-4.7$ \\
\object{UGC~9052} & $99.2\pm6.8$\phn & \nodata & \nodata & \nodata & $15.4\pm1.5$ & \nodata & $10.3\pm0.2$ & $28\pm6$ ($0\pm44$) & $137\pm24$\tablenotemark{$^\dagger$} & $6.4\pm0.4$ & $\approx$0 & $-3.4$ \\
\object{UGC~9340} & $62.9\pm6.6$\phn & \ion{H}{2} & 5 & 0 & $19.6\pm2.4$ & \nodata & $9.9\pm0.2$ & $23\pm10$ ($21\pm14$) & $144\pm60$ & $6.0\pm0.8$ & 9.9 & $-1.3$ \\
\object{UGC~9722} & $95.9\pm8.0$\phn & \nodata & \nodata & \nodata & $6.2\pm4.7$ & \nodata & $10.0\pm0.2$ & $29\pm6$ ($29\pm12$) & $113\pm23$\tablenotemark{$^\dagger$} & $7.0\pm0.6$ & 0.1 & $-3.2$ \\
\object{UGC~10020} & $31.5\pm6.8$\phn & \ion{H}{2} & \nodata & \nodata & $19.2\pm3.5$ & \nodata & $9.6\pm0.2$ & $19\pm2$ ($12\pm30$) & \phn$77\pm7$\phn & $5.0\pm0.5$ & 52.0 & 0.1 \\
\object{UGC~10146}	& $102.2\pm6.1$\phn & \nodata & \nodata & \nodata & $14.2\pm2.9$ & \nodata & $10.1\pm0.2$ & $34\pm1$ ($12\pm36$) & $127\pm6$\phn & $6.4\pm0.4$ & $\approx$0 & $-3.6$ \\
\object{UGC~10440}	& $59.1\pm6.4$\phn & \nodata & \nodata 	& \nodata 	& $34.5\pm1.9$ & \nodata & $9.6\pm0.2$ & $33\pm7$ ($17\pm21$) & \phn$68\pm13$ & $2.9\pm0.5$ & $\approx$100 & 4.4 \\
\object{UGC~10831}	& $102.2\pm6.1$\phn & \nodata & \nodata & \nodata & $20.2\pm3.5$ & \nodata & $10.4\pm0.1$ & $41\pm2$ ($27\pm13$) & $275\pm18$ & $7.1\pm0.5$ & $\approx$0 & $-4.5$ \\
\object{UGC~10922}	& $113.5\pm6.4$\phn & \nodata & \nodata & \nodata & $16.8\pm0.8$ & \nodata & $10.2\pm0.2$ & $25\pm5$ ($21\pm21$) & $118\pm24$ & $6.0\pm0.4$ & 1.1 & $-2.3$ \\
\object{UGC~11029}	& $40.4\pm5.9$\phn	& \nodata & \nodata & \nodata & $24.3\pm4.8$ & \nodata & $9.6\pm0.2$ & $39\pm0$ ($0\pm44$) & $123\pm5$\phn & $5.2\pm0.6$ & 36.5 & $-0.3$ \\
\object{UGC~11113}	& $31.4\pm6.1$\phn & \nodata & \nodata & \nodata & $19.7\pm2.9$ & \nodata & $9.4\pm0.2$ & $54\pm1$ ($24\pm15$) & \phn$55\pm2$\phn & $4.3\pm0.4$ & 95.2 & 1.7 \\
\object{UGC~11386}	& $102.5\pm6.6$\phn & \nodata & \nodata & \nodata & $12.6\pm2.7$ & \nodata & $10.2\pm0.2$ & $27\pm3$ ($21\pm41$) & $146\pm19$ & $6.8\pm0.4$ & $\approx$0 & $-4.2$ \\
\object{UGC~11515}	& $41.3\pm30.6$ & \nodata & \nodata & \nodata & $25.4\pm0.3$ & \nodata & $10.1\pm0.7$ & $36\pm2$ ($0\pm44$) & $109\pm11$ & $4.9\pm0.3$ & 66.6 & 0.4 \\
\object{UGC~11556}	& $71.0\pm6.2$\phn & \nodata & \nodata & \nodata & $23.8\pm3.4$ & \nodata & $9.5\pm0.2$ & $30\pm0$ ($0\pm53$) & \phn$79\pm3$\phn & $4.5\pm0.5$ & 85.3 & 1.1 \\
\object{UGC~11653}	& $47.6\pm5.5$\phn & \nodata & \nodata & \nodata & $17.5\pm2.2$ & \nodata & $9.7\pm0.2$ & $42\pm3$ ($17\pm25$) & $108\pm7$\phn & $5.8\pm0.3$ & 1.5 & $-2.2$ \\
\object{UGC~11699}	& $58.4\pm5.9$\phn & \nodata & \nodata & \nodata & $15.0\pm1.3$ & \nodata & $9.8\pm0.2$ & $27\pm4$ ($21\pm17$) & $103\pm14$ & $5.9\pm0.4$ & 0.4 & $-2.7$ \\
\object{UGC~11728}	& $108.1\pm6.6$\phn & \nodata & \nodata & \nodata & $11.5\pm3.2$ & \nodata & $10.4\pm0.1$ & $22\pm3$ ($29\pm14$) & $103\pm13$ & $6.3\pm0.5$ & 0.2 & $-2.9$ \\
\object{UGC~11992}	& $42.8\pm5.8$\phn & \nodata & \nodata & \nodata & $9.2\pm2.0$ & \nodata & $9.4\pm0.2$ & $44\pm2$ ($24\pm35$) & $103\pm4$\phn & $6.5\pm0.3$ & $\approx$0 & $-5.0$ \\
\object{UGC~12008}	& $104.9\pm6.8$\phn & \nodata & \nodata & \nodata & $11.0\pm4.5$ & \nodata & $10.3\pm0.2$ & $31\pm3$ ($27\pm16$) & $210\pm22$ & $7.6\pm0.5$ & $\approx$0 & $-4.8$ \\
\object{UGC~12015}	& $100.4\pm6.4$\phn & \nodata & \nodata & \nodata & $12.0\pm4.7$ & \nodata & $10.2\pm0.2$ & $40\pm2$ ($27\pm10$) & \phn$99\pm7$\phn	& $6.2\pm0.5$ & 1.4 & $-2.2$ \\
\object{UGC~12176}	& $123.5\pm6.7$\phn & \nodata & \nodata & \nodata & $15.4\pm3.9$ & \nodata & $10.5\pm0.2$ & $41\pm3$ ($27\pm13$) & \phn$52\pm3$\phn & $4.7\pm0.5$ & 73.1 & 0.6 \\
\object{UGC~12184}	& $51.2\pm6.1$\phn & \nodata & \nodata & \nodata & $16.0\pm3.0$ & \nodata & $9.5\pm0.1$ & $19\pm3$ ($0\pm44$) & \phn$90\pm14$ & $5.6\pm0.5$ & 9.8 & $-1.3$ \\
\object{UGC~12289}	& $135.8\pm6.6$\phn & \nodata & \nodata & \nodata & $24.0\pm4.7$ & \nodata & $10.5\pm0.2$ & $29\pm8$ ($17\pm21$) & $201\pm53$ & $6.1\pm0.7$ & 7.1 & $-1.5$ \\
\object{UGC~12685}	& $67.0\pm6.3$\phn & \nodata & \nodata & \nodata & $20.7\pm3.9$ & \nodata & $9.8\pm0.2$ & $28\pm10$ ($17\pm17$) & $125\pm40$ & $5.7\pm0.7$ & 19.0 & $-0.9$ \\
\enddata
\tablecomments{
\textbf{Column~(1):} galaxy name. 
\textbf{Column~(2):} luminosity distance (in Mpc) from HyperLeda \citep{Makarov:2014}.
Galaxies closer than 200\,Mpc have been adjusted according to the Cosmicflows-3 Distance--Velocity Calculator \citep{Kourkchi:2020}, available at the Extragalactic Distance Database (\url{http://edd.ifa.hawaii.edu}).
Distances (d) less than 38\,Mpc have computed expectation distances based on the smoothed velocity field from the Numerical Action Methods model of \citet{Shaya:2017}.
$38\,{\rm Mpc}<d<200\,{\rm Mpc}$ have computed expectation distances based on the smoothed velocity field from the linear density field model of \citet{Graziani:2019}.
Redshift-dependent distances are calibrated to the \citet{Planck:2020} cosmographic parameters.
\textbf{Column~(3):} the \citetalias{Baldwin:1981} [\ion{N}{2}]/H$\alpha$ versus [\ion{O}{3}]/H$\beta$ standard optical diagnostic classification.
``\ion{H}{2}'' = \ion{H}{2}-region-like galaxy, ``AGN'' = active galactic nucleus, and ``Comp'' = composite galaxy (likely to contain a metal-rich stellar population plus an AGN). 
\textbf{Column~(4):} Chandra X-ray Observatory (CXO) exposure time (in ks) of the galaxy's nucleus. 
\textbf{Column~(5):} the number of X-ray photons emanating from the galaxy's nucleus detected by the CXO. 
\textbf{Column~(6):} absolute value of the \emph{face-on} spiral-arm pitch angle (in degrees), derived primarily from Pan-STARRS1 imaging, and measured by the \href{https://github.com/bendavis007/2DFFT}{\textcolor{linkcolor}{\texttt{2DFFT}}}, \href{http://sparcfire.ics.uci.edu/}{\textcolor{linkcolor}{\texttt{SpArcFiRe}}}, and/or \href{https://github.com/DeannaShields/Spirality}{\textcolor{linkcolor}{\texttt{Spirality}}} software packages.
\textbf{Column~(7):} central stellar velocity dispersion (in km\,s$^{-1}$) from HyperLeda.
\textbf{Column~(8):} logarithm of the total galaxy stellar mass (in $\mathrm{M}_\sun$); we use the intrinsic $K$-band apparent magnitudes from HyperLeda, then converted to luminosity by correcting for distance, compensating for surface brightness dimming \citep{Tolman:1930,Tolman:1934}, and applying the solar absolute magnitude from \citet{Willmer:2018}, which is multiplied by the mass-to-light ratio from \citet[][equation~8]{Davis:2019b}.
\textbf{Column~(9):} inclination angle (in degrees), determined via our pitch angle measurement process. The parenthetical values are the inclination angles according to the \citetalias{RC3}.
\textbf{Column~(10):} physical maximum velocity rotation (in km\,s$^{-1}$) corrected for inclination, from HyperLeda. 
\textbf{Column~(11):} predicted black hole mass from the planar relation of \citet{Davis:2023}, \textit{i.e.}, combining Columns~(6) and (10).
\textbf{Column~(12):} probability that the central black hole is an IMBH, \textit{i.e.}, $P(\mathcal{M}_\bullet\leq5)$ \citep[via][equation~7]{Davis:2020}.
\textbf{Column~(13):} number of standard deviations below $10^5\,\mathrm{M}_\sun$, \textit{i.e.}, $n\,\sigma=(5-\mathcal{M}_\bullet)/\delta\mathcal{M}_\bullet$.
}
\tablenotetext{\ast}{Obtained from the NASA-Sloan Atlas (\url{http://www.nsatlas.org}).
The $1\farcs5$-radius-aperture-based $\sigma_0$ value was normalized to match the $0.595\,h^{-1}\,$kpc-radius-aperture used in the homogenized system of HyperLeda via the prescriptions of \citet{Jorgensen:1995}.}
\tablenotetext{\dagger}{Velocity is estimated via the Tully-Fisher relation \citep{Tully:1977} as refined by \citet{Tiley:2019}.}
\end{deluxetable*}

\section{Predicting Black Hole Masses}\label{sec:masses}

\subsection{The $M_\bullet$--$\phi$ Relation}

The $M_\bullet$--$\phi$ relation is the most natural black hole mass scaling relation for application to spiral galaxies.
Furthermore, it exhibits the lowest level of intrinsic scatter \citep[$\epsilon=0.33\pm0.08$\,dex from][]{Davis:2017} of any single black hole mass scaling relation for spiral galaxies.
The logarithmic spiral pitch angle is the most-widely adopted metric for quantifying the geometric shape of spiral arms in disk galaxies.
The $M_\bullet$--$\phi$ relation \citep{Seigar:2008,Berrier:2013,Davis:2017} can be used to predict central black holes in spiral galaxies and identify IMBH candidates \citep{Graham:2019,Treuthardt:2019}.
The $M_\bullet$--$\phi$ relation is uniquely applicable to only spiral galaxies.
Whereas most other black hole mass scaling relations may be employed for any morphological type of galaxy, such scaling relations are usually less accurate when derived from spiral galaxies alone.
For example, \citet{Sahu:2019} found that the $M_\bullet$--$\sigma_0$ relation for late-type galaxies exhibits an intrinsic scatter that is 0.25\,dex higher than the $M_\bullet$--$\sigma_0$ relation for early-type galaxies.

We have measured $\phi$ for all 85 galaxies in our sample (Column~6 of Table~\ref{tab:sample}).
We obtained $g$, $r$, $i$, $z$, and $y$ images from Pan-STARRS1\footnote{\url{https://ps1images.stsci.edu/cgi-bin/ps1cutouts}} \citep{Chambers:2016}, and the image that best highlighted the spiral structure was adopted for the pitch angle measurement.
Sloan Digital Sky Survey (SDSS) or Galaxy Evolution Explorer (GALEX) images were also consulted and used if the volute structure was better resolved.
Galaxy images were first deprojected to a face-on orientation and then analyzed by a combination of software packages, including \href{https://github.com/bendavis007/2DFFT}{\textcolor{linkcolor}{\texttt{2DFFT}}} \citep{Davis:2012,2dfft}, \href{http://sparcfire.ics.uci.edu/}{\textcolor{linkcolor}{\texttt{SpArcFiRe}}} \citep{sparcfire}, and/or \href{https://github.com/DeannaShields/Spirality}{\textcolor{linkcolor}{\texttt{Spirality}}} \citep{spirality,Shields:2022}.\footnote{
See \S2.1 from \citet{Davis:2020} for further details.}

\subsection{The $M_\bullet$--$v_\mathrm{max}$ Relation}

For most of our sample ($75/85\approx88\%$), we were able to obtain the physical maximum rotational velocities ($v_\mathrm{max}\equiv v_{\rm rot}$) from HyperLeda \citep{Makarov:2014}.
As for the remaining ten galaxies (see Table~\ref{tab:sample}, Column~10), we estimated the rotational velocities by applying the Tully-Fisher relation \citep{Tully:1977} as refined by \citet{Tiley:2019}.
The HyperLeda velocities are derived from the original line-of-sight velocities (from both 21-cm line widths and/or rotation curves) and subsequently corrected for inclination.
Although we used the inclination angles derived from the axial ratios taken from the \citetalias{RC3}\footnote{See the parenthetical values from Column~9 of Table~\ref{tab:sample}.} to initially restrict our sample to only face-on ($\theta\leq30\degr$) galaxies, we elected to use our inclination angles that were derived during the process of measuring pitch angle.
We find an unremarkable agreement between our measured inclinations and those from the \citetalias{RC3}, primarily because the latter have high levels of uncertainty.
Indeed, the median agreement between the inclinations is $0.53\,\sigma$ (i.e., the joint number of standard deviations at which both observations agree) or approximately 60\% mutual agreement.
Most importantly, it was necessary to use our measured inclinations because 19 of the galaxies from the \citetalias{RC3} were listed as essentially face on (i.e., $\theta\approx0$), thus making rotational velocity measurements impossible, which is incompatible with the existence of line-of-sight velocities from HyperLeda.

In fact, the act of determining inclination angles via the process of measuring pitch angles and/or Fourier transforms of spiral galaxy images has shown itself to be an accurate alternative to traditional inclination measurements via isophotal analysis \citep[\textit{e.g.},][]{Grosbol:1985,Ma:2001,Gomez:2004,Poltorak:2007,Fridman:2010}.
Thus, reversing the procedure can yield precise inclination angle measurements of galaxies by adjusting the inclination angle until a galaxy's spiral structure is closely described by pure logarithmic spirals with $\phi$ such that the growth of the spirals (radius $R$ as a function of galactocentric azimuth $\varphi$) is monotonic ($\dv*{R}{\varphi}>0$).
As a result, some of our galaxies were subsequently found to have inclinations $>$$30\degr$, but we obtained more precise measurements.
Our inclination estimates have a mean uncertainty of $3\fdg4$, compared to $18\fdg5$ for those from the \citetalias{RC3} axial ratios.\footnote{
From their study of uncertainties in the projection parameters of spiral galaxies, \citet{Barnes:2003} found systematic uncertainties of $\approx$$4\degr$ in inclination derived from photometry due to the presence of nonaxisymmetric structures.}
The $M_\bullet$--$v_\mathrm{max}$ and $M_\bullet$--$\phi$ relations are both determined from the same sample of galaxies, but the former exhibits a higher level of intrinsic scatter, $\epsilon=0.45$\,dex \citep[equation~10 from][]{Davis:2019}.

\subsection{A Planar Black Hole Mass Scaling Relation}\label{sec:FP}

Ever since the \citet{Magorrian:1998} relation came onto the scene, the breadth and variety of black hole mass scaling relations has grown at a seemingly exponential rate in astrophysical literature.
In our larger study \citep{Jin:Davis:2023}, we endeavored to use modern machine learning methods to try and identify higher-dimensional black hole mass scaling relations (\textit{i.e.}, $M_\bullet$ plus \emph{two or more} galactic parameters) that are more accurate predictors of black hole mass than some of the aforementioned two-dimensional black hole mass scaling relations (\textit{i.e.}, $M_\bullet$ plus \emph{one} galactic parameter).\footnote{Machine learning has also been utilized recently to try and identify globular clusters that host IMBHs \citep{Pasquato:2023}.}
For that study, we utilized a subfield of machine learning called symbolic regression \citep{Cranmer:2023} to search for the best mathematical expressions to fit our dataset of dynamically-measured SMBH masses and their host galaxy parameters.
The parent study of \citet{Jin:Davis:2023} identified an ideal (\textit{i.e.}, optimally precise and simple) relationship among $M_\bullet$, $\phi$, and $v_\mathrm{max}$ for spiral galaxies.
We presented the details behind this trivariate relationship separately in \citet{Davis:2023}.
 
\begin{figure}
\includegraphics[clip=true, trim= 19mm 17mm 5mm 21mm, width=\columnwidth]{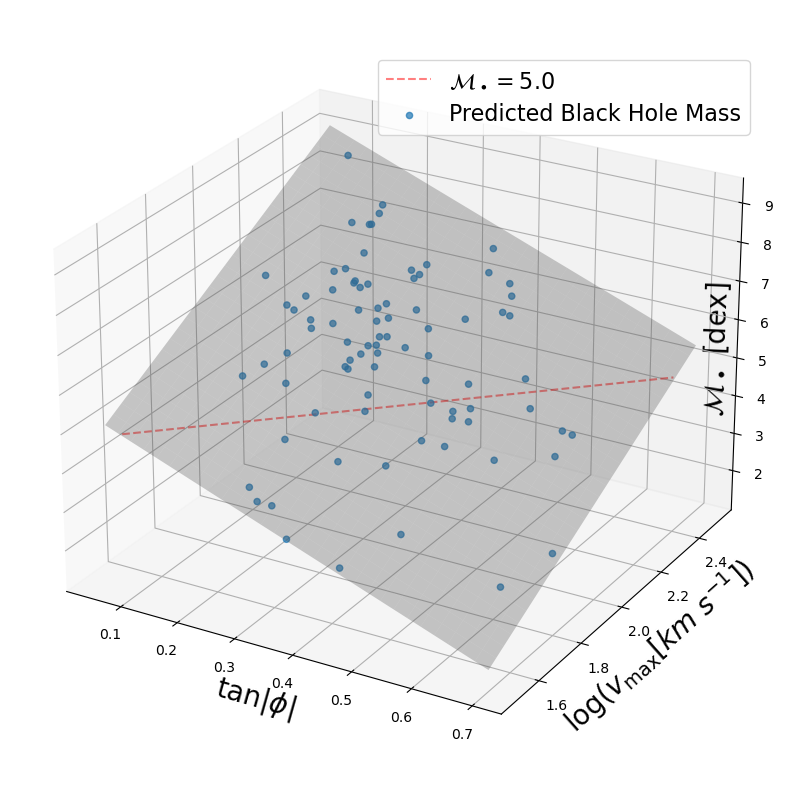}
\caption{
Here, we have reproduced the three-dimensional plot of the planar $M_\bullet$--$\phi$--$v_\mathrm{max}$ relationship (Equation~\ref{eqn:FP}) from \citet[][figure~2]{Davis:2023}.
Onto the surface (\textcolor{lightgray}{$\blacklozenge$}), we show the locations of all 85 of our galaxies from this work (\textcolor{MATLAB_blue}{$\bullet$}) and a demarcation line (\textcolor{MATLAB_red}{{\hdashrule[0.35ex]{8mm}{1pt}{1mm}}}) showing the boundary between supermassive and intermediate-mass black holes at $\mathcal{M}_\bullet=5.0$.
As shown, 23/85 (27\%) of the galaxies lie below the dividing line, representing our IMBH-candidate host galaxies.
Moreover, this enhanced plot illustrates that our galaxies are dispersed over the area of the plane, demonstrating a lack of degeneracy between the parameters by the apparent embedding of the two-dimensional manifold (\textit{i.e.}, surface) in three-dimensional space. 
For an animation of this plot, see the following link, \url{http://surl.li/kkjeu}.
}
\label{fig:FP}
\end{figure}

From \citet{Davis:2023}, the equation for the $M_\bullet$--$\phi$--$v_\mathrm{max}$ relationship is:
\begin{equation}
\mathcal{M}_\bullet=\alpha(\tan|\phi|-0.24)+\beta\log\left(\frac{v_\mathrm{max}}{211\,\mathrm{km\,s}^{-1}}\right)+\gamma,
\label{eqn:FP}
\end{equation}
with $\alpha=-5.58\pm0.06$, $\beta=3.96\pm0.06$, $\gamma=7.33\pm0.05$, and $\epsilon=0.22\pm0.06$\,dex in the $\mathcal{M}_\bullet$-direction, with parameters identified by \href{https://github.com/MilesCranmer/PySR/tree/v0.12.3}{\textcolor{linkcolor}{\texttt{PySR}}} \citep{Cranmer:2023} and refined via \href{https://github.com/CullanHowlett/HyperFit}{\textcolor{linkcolor}{\texttt{Hyper-Fit}}} \citep{Robotham:2015,Robotham:2016}.
We utilize Equation~\ref{eqn:FP} to predict black hole masses for all of our sample (see Column~11 of Table~\ref{tab:sample}).
Figure~\ref{fig:FP} illustrates a plot of the plane and the location of all 85 of our galaxies on the plane.
The orientation of the plane conforms with expectations---\textbf{big} black holes reside in galaxies that have \underline{tightly} wound spiral arms \emph{and} whose disks are \underline{rapidly} rotating, and \textbf{small} black holes are found in galaxies that have \underline{loosely} wound spiral arms \emph{and} whose disk are \underline{slowly} rotating.
The combination of these parameters is not surprising; \citet{Sarkar:2023} find that later morphological types are primarily flocculent (in contrast to grand-design) galaxies with more loosely wound spiral arms and slower rotational velocities.
For additional analyses and discussions, see \citet{Davis:2023} regarding the planar relation for spiral galaxies and \citet{Jin:Davis:2023} for higher-dimensional relations featuring all galaxy types.

We present a couple of metrics to help codify our best IMBH candidate host galaxies in the rightmost columns of Table~\ref{tab:sample}.
Column~12 displays the probability that the central black hole is an IMBH, \textit{i.e.}, $P(\mathcal{M}_\bullet\leq5)$ \citep[via][equation~7]{Davis:2020}.
Column~13 provides the number of standard deviations below $10^5\,\mathrm{M}_\sun$, \textit{i.e.}, $n\,\sigma=(5-\mathcal{M}_\bullet)/\delta\mathcal{M}_\bullet$, which is equivalent to $P(\mathcal{M}_\bullet\leq5)$, \textit{e.g.}, $P(\mathcal{M}_\bullet\leq5) = 68.3\% \equiv 1.0\,\sigma$.
Together, we utilize these quality/qualifying parameters to categorize our sample into striations of likeliness that a galaxy is expected to harbor an IMBH.

Inclination ($\theta$) is the one common source of error that affects both independent variables ($|\phi|$ and $v_\mathrm{max}$) in our study.
Because of the small-angle approximation, an error in the inclination of a galaxy when it is close to face-on ($0\degr$) is more significant than an equally-sized error when a galaxy is close to edge-on ($90\degr$).
This heteroscedasticity negatively affects the calculation of the intrinsic $v_\mathrm{max}$ from the observed line-of-sight velocity ($v_{\rm LOS}$), because $v_\mathrm{max}\equiv v_{\rm LOS}\csc{\theta}$.
However, the opposite effect is applicable to the measurement of pitch angles.
Because a galaxy must be artificially projected into a face-on orientation to measure pitch angle, galaxies that are already close to a face-on orientation require minimal modification.
Specifically, the minor-axis length ($b$) of a galaxy is stretched to equal its major-axis length ($a$), \textit{i.e.}, $a\equiv b\sec{\theta}$.

\section{Results}\label{sec:results}

\subsection{Typical Central Black Hole in an Sd Galaxy}\label{sec:SdBH}

\begin{figure}
\includegraphics[clip=true, trim= 0mm 0mm 0mm 0mm, width=\columnwidth]{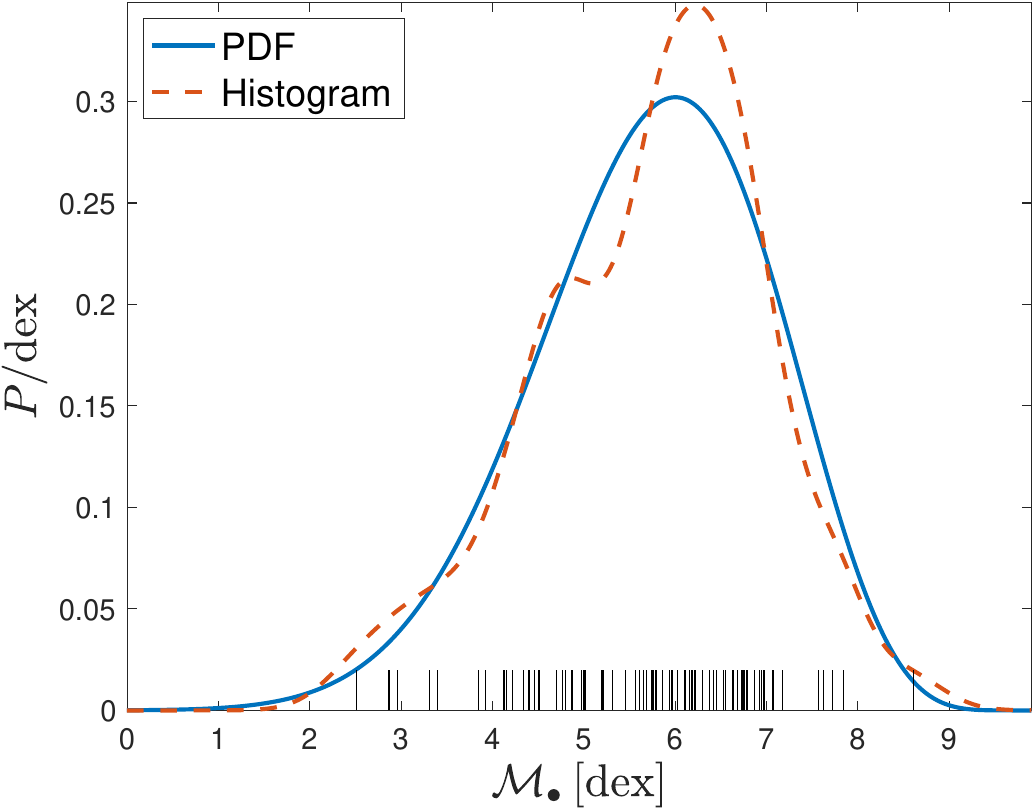}
\caption{
The distribution of black hole mass estimates ($\mathcal{M}_\bullet$) for the 85 Sd galaxies in our sample (from Column~11 of Table~\ref{tab:sample}) is shown as a rug plot along the bottom axis and has a mean $\mathcal{M}_\bullet$ of $5.73\pm1.30$ and a median of $5.89\pm0.86$.
The smoothed histogram (\textcolor{MATLAB_red}{{\hdashrule[0.35ex]{8mm}{1pt}{1mm}}}) is generated from the summation of all 85 mass estimates.
The fitted PDF (\textcolor{MATLAB_blue}{{\hdashrule[0.35ex]{8mm}{0.4mm}{}}}) for the Sd galaxies peaks at $\mathcal{M}_\bullet=6.00\pm0.14$ (its mode).
Here, the integrated probability under the PDF at $\mathcal{M}_\bullet\leq5$ is 27.7\%.
}
\label{fig:all}
\end{figure}

In Figure~\ref{fig:all}, we produce a histogram by summation of all 85 black hole mass estimates (determined by Equation~\ref{eqn:FP}) for our sample of 85 Sd galaxies.
We then fit a skew-kurtotic-normal distribution to the histogram.
In doing so, we find that the typical Sd galaxy hosts a black hole with $\mathcal{M}_\bullet=6.00\pm0.14$, with $P(\mathcal{M}_\bullet\leq5)=27.7\%$.
Expressed differently, this most probable mass of $(1.01\pm0.33)\times10^6\,\mathrm{M}_\sun$ is $25.1\%\pm8.2\%$ the mass of Sgr~A$^\ast$ \citep{Boehle:2016}.

Nominally, we expect that 1/3.61 of Sd galaxies harbor an IMBH.
The \citetalias{RC3} provides morphological classifications for 17,801 diameter-selected galaxies, of which 787 are classified as Sd, \textit{i.e.}, $4.42\%$.
A similar value of $3.9\%$ (34/867) was found for Sd types by the diameter-selected sample of \citet{Lacerda:2020}.
However, the magnitude-limited sample of the Carnegie-Irvine Galaxy Survey \citep{CGS}, plus the Milky Way, shows only 9/606 galaxies (1.49\%) are type Sd, whereas its magnitude- and volume-limited subsample shows only 1/208 galaxies (0.5\%) are type Sd \citep{Davis:2014,Mutlu-Pakdil:2016}.
The EFIGI catalogue \citep{Baillard:2011} is a subset of 4,458 galaxies from the \citetalias{RC3}, but no longer considers peculiar galaxies or galaxies with special features as belonging to separate stages.
Thus, they classify an even higher fraction (285/4,45$8=6.39\%$) of galaxies as $T=7$ \citep{Lapparent:2011}.

It is worth noting that there is an even higher fraction (\textit{i.e.}, almost double) of Sd galaxies in the nearby Universe; there are 39 Sd galaxies out of 454 galaxies with a classification (8.6\%) in the \citetalias{RC3} at $cz<1$,000\,km\,s$^{-1}$ ($z<0.003$; luminosity distances out to $d=15$\,Mpc).
This could indicate a real local excess or maybe a bias against Sd identification in more distant galaxies.
Given this frequency of Sd galaxies in our local Universe, we would, therefore, expect that $>$218/17,801 or $>$1.22\%\footnote{
This is a lower limit because we are not accounting here for other potential hosts of IMBHs (\textit{e.g.}, dwarf galaxies and globular clusters) or their frequency (albeit, to a lesser extent) in earlier type spirals.} of bright galaxies ($B_{\rm T}\lesssim15.5$\,mag) in our local Universe host an IMBH, \textit{i.e.}, the ``occupation fraction'' \citep{Zhang:2009,Miller:2015,Gallo:2019}.
Using the cosmographic parameters from \citet{Planck:2020}, this gives a comoving volume of $4.40\times10^{7}$\,Mpc$^{3}$ for the \citetalias{RC3} extent and a lower limit for the number density of central IMBHs in our local Universe $>$$4.96\times10^{-6}$\,Mpc$^{-3}$.

\subsection{Additional Observational Data}\label{sec:add}

\subsubsection{\citetalias{Baldwin:1981} Classifications}

We consulted the Reference Catalog of galaxy SEDs \citep[RCSED;][]{Chilingarian:2017} to obtain nuclear classifications for each galaxy.
The RCSED database utilizes the Baldwin-Phillips-Terlevich \citep[hereafter \citetalias{Baldwin:1981};][]{Baldwin:1981} [\ion{N}{2}]/H$\alpha$ versus [\ion{O}{3}]/H$\beta$ standard optical diagnostic classification, according to \citet{Kewley:2006}.
As \citet{Birchall:2021} point out, the \citetalias{Baldwin:1981} diagnostic clearly fails at reliably identifying sources as AGNs in lower mass galaxies.
Moreover, emission line flux ratio diagnostics can fail to identify entire subpopulations of AGNs when applied to single-fiber optical spectra \citep{Comerford:2022}.

Other diagnostics, such as \ion{He}{2}, may be better suited to detecting ionizing photons from AGNs in low-metallicity, star-forming dwarf galaxies \citep{Umeda:2022,Tozzi:2023}.
As such, galaxies are classified as containing an \ion{H}{2}-region, AGN, or both (composite).
Only one galaxy (UGC~8171) is classified as having a \textit{bona fide} AGN,\footnote{For comparison, \citet{Lacerda:2020} found that none of the 34 Sd galaxies in their sample possess an AGN.} and another is a composite (UGC~4622).
Coincidentally, both galaxies are predicted to have similar black hole masses (UGC~4622: $\mathcal{M}_\bullet=6.16\pm0.46$, with $P(\mathcal{M}_\bullet\leq5)=0.6\%$; UGC~8171: $\mathcal{M}_\bullet=6.55\pm0.39$, with $P(\mathcal{M}_\bullet\leq5)\approx0\%$), which are strongly inconsistent with being IMBHs.

Fourteen galaxies (including the composite UGC~4622) are classified as \ion{H}{2}-region-like galaxies.
The sample peak probability (mode) black hole mass for these 14 galaxies is $\mathcal{M}_\bullet=5.14\pm0.35$, with $P(\mathcal{M}_\bullet\leq5)=36.0\%$.
Therefore, these 14 \ion{H}{2}-region-like galaxies have a slightly higher probability of containing central IMBHs than the typical Sd galaxy.
A possible explanation for this difference could emerge from varied gas fractions between these subpopulations of galaxies (see a further discussion in \S\ref{sec:ave}).

\subsubsection{Nuclear X-ray Point Sources}

We checked the CXO archive for observations of all galaxies in our sample.
The CXO has observed eight of our galaxies; however, only one galaxy (NGC 2500) exhibited a measurable flux of X-ray photons emanating from its nucleus.
With $\mathcal{M}_\bullet=5.57\pm0.70$ and $P(\mathcal{M}_\bullet\leq5)=20.7\%$, NGC~2500 has a slightly higher probability of possessing an IMBH than the typical Sd galaxy.
The seven galaxies without a measurable nuclear X-ray source have a sample mode black hole mass of $\mathcal{M}_\bullet=5.79\pm0.41$, \emph{but} $P(\mathcal{M}_\bullet\leq5)=41.4\%$ with a strong tail towards lower masses.
Thus, these seven galaxies, where X-ray sources have been sought (but not found), are exactly \emph{twice} as likely to harbor an IMBH in their nuclei (as compared to NGC~2500).
Put another way, finding a nuclear X-ray source implies that a galaxy is twice as likely to harbor an SMBH, instead of an IMBH.
Therefore, non-detection of nuclear X-ray sources could be a useful diagnostic for identifying IMBH host galaxy candidates.

This paltry harvest of nuclear X-ray sources for our sample is not unexpected, because we deliberately selected intrinsically faint Sd galaxies.
With average black holes masses less than a million solar masses and Eddington ratios of $\sim$$10^{-6}$, the typical Sd galaxy will likely have weak or no AGN signature.
Also, Sd galaxies presumably lack one alleged mechanism behind AGN triggering, \textit{i.e.}, major mergers.
Models suggest that major mergers may establish galaxy-wide, gravitationally-induced torques that drive gas toward a galactic center, which may set off an AGN \citep[\textit{e.g.},][]{Hopkins:2006}.
However, the jury is decidedly hung when it comes to reaching a verdict about whether major mergers are relevant \citep[\textit{e.g.},][]{Koss:2010,Ellison:2011,Hong:2015,Weston:2017,Goulding:2018,Gao:2020,Toledo:2023,Breiding:2023} or not \citep[\textit{e.g.},][]{Cisternas:2011,Kocevski:2012,Karouzos:2014,Villforth:2019,Lambrides:2021} in triggering AGN.
Optical emission line diagnostic tests for activity and/or X-ray emission from nuclei are usually reliable indications for the presence of an accreting SMBH.
However, the lack of such indicators from the centers of Sd galaxies is not evidence for the lack of black holes, which may be in a quiescent phase of their duty cycles.

The occupation fraction of galaxies with nuclear X-ray emission can be quite different, depending on the sample selection.
For example, \citet{Williams:2022} found X-ray emission coincident within $2^{\prime\prime}$ of their optical galaxy centers in the majority of their galaxies ($150/213=70.4\%$).
Whereas, \citet{Nwaokoro:2021} found X-ray sources within $3^{\prime\prime}$ of their host galaxy centers in only a tiny fraction of their sample ($7/1200=5.83$\textperthousand).
The difference being that the former sample \citep{Williams:2022} was constructed as a statistically complete sample of nearby galaxies, while the latter sample \citep{Nwaokoro:2021} is a sample of only low-mass galaxies ($M_{\rm gal,\star}\lesssim10^{9.5}\,\mathrm{M}_\sun$).
Moreover, \citet{Ohlson:2023} found that in low-mass late-type galaxies X-ray fractions are lower and positional offsets of X-ray detections from their galactic centers are higher (possibly due to increased astrometric uncertainty).
Thus, in general, exceedingly few X-ray sources exist in low-mass galaxies.

Moreover, an IMBH might not reside at the center of a galaxy.
Due to their diminished mass, IMBHs traverse a difficult path sinking to the centers of their host galaxies and are more prone to wander away \citep{Binggeli:2000,Bellovary:2019,Bellovary:2021,Barrows:2019,Pfister:2019,Reines:2020,Mezcua:2020,Ma:2021,Ricarte:2021,Cintio:2023,Partmann:2023} as compared to their more massive SMBH cousins.
However, \citet{Chu:2022} point out that off-center SMBHs are common in brightest cluster galaxies due to numerous galaxy mergers in their history, in which, an SMBH can become significantly kicked out of the galactic center via dynamical interactions.
Such offset SMBHs could potentially be detected via distortions that they cause in gravitational lens galaxies \citep{Perera:2023,Giani:2023}.
In particular, perhaps the best (and most studied) IMBH candidate is HLX-1, which is 3.7\,kpc from the center of \object[ESO 243-49]{ESO~243-49}:  \citep{Farrell:2009,Farrell:2012,Soria:2010,Soria:2011,Soria:2012,Soria:2013,Webb:2010,Webb:2012,Webb:2017}.\footnote{
See also, the much-discussed ultra-luminous source from a possible IMBH in \object{Messier~82} \citep{Patruno:2006,Muxlow:2010,Joseph:2011,Pasham:2014}.
}
However, \citet{Weller:2022b} and \citet{DiMatteo:2023} remark about the largely-missing population of wandering IMBHs, which rarely reveal themselves in the hyper-luminous X-ray sources (HLXs) regime.
IMBHs may also be associated with the merging remnants of dwarf galaxies onto larger galaxies \citep{Webb:2010,Webb:2017,Farrell:2012,Mapelli:2012b,Soria:2013,Mezcua:2015,Kim:2015,Kim:2017,Kim:2020,Graham:2021b}.

\subsection{IMBH Targets of Interest}\label{sec:targets}

We define a target of interest to have $P(\mathcal{M}_\bullet\leq5)>50\%$ ($n\,\sigma>0$).
Looking down Table~\ref{tab:sample} Column~12, we identify 23 targets of interest (points plotted below the dashed red line in Figure~\ref{fig:FP}).
Indeed, this fraction of galaxies (23/85=27\%) is comparable with the $P(\mathcal{M}_\bullet\leq5)=27.7\%$ value we obtain from Figure~\ref{fig:all}.
For these 23 galaxies, we wager that the odds are favorable that they each host a central IMBH.
Among these targets, UGC~1544 is our most likely example to host an IMBH, with an expected mass of only $323\pm293\,\mathrm{M}_\sun$.
Moreover, these 23 galaxies are all relatively nearby; their mean distance is $55.2\pm29.2$\,Mpc, median distance is $52.5\pm18.2$\,Mpc, and a peak probability distance of $42.1\pm6.1$\,Mpc.
Thus, these galaxies pose appealing targets for further study.

\subsubsection{Targets with Additional Consideration}

\textbf{UGC~3826} possesses an NSC with a mass of $\mathcal{M}_{\rm NSC,\star}=6.04\pm0.09$ \citep{Georgiev:2014,Georgiev:2016}.
By application of the $M_\bullet$--$M_{\rm NSC,\star}$ relation \citep[][equation~9]{Graham:2019c}, we predict a black hole mass of $\mathcal{M}_\bullet=3.53\pm0.95$, with $P(\mathcal{M}_\bullet\leq5)=93.9\%$.
Additionally, we can apply the $M_\bullet$--$M_{\rm gal,\star}$ relation \citep[][equation~3]{Davis:2018} to the total stellar mass of UGC~3826 ($\mathcal{M}_{\rm gal,\star}=9.33\pm0.27$) and predict a black hole mass of $\mathcal{M}_\bullet=2.75\pm1.30$, with $P(\mathcal{M}_\bullet\leq5)=95.8\%$.
As we can see, these black hole mass predictions are very consistent with the value we found from Equation~\ref{eqn:FP}, $\mathcal{M}_\bullet=3.40\pm0.60$, with $P(\mathcal{M}_\bullet\leq5)=99.6\%$.

Additionally, we identified two of our galaxies with black hole mass estimates via the $M_\bullet$--$\mathcal{C}_\mathrm{FUV,tot}$ and $M_\bullet$--$\mathcal{C}_\mathrm{NUV,tot}$ relations of \citet{Dullo:2020}, where $\mathcal{C}_\mathrm{FUV,tot}$ is the total FUV$-[3.6\,\micron]$ color and $\mathcal{C}_\mathrm{NUV,tot}$ is the total NUV$-[3.6\,\micron]$ color.
As expected, the $M_\bullet$--$\mathcal{C}$ relations from \citet{Dullo:2020} are such that more massive black holes reside in redder galaxies.
Similarly, \citet{Baker:2023} found that more massive black holes reside in galaxies with higher metallicities.
For \textbf{UGC~8153}, \citet{Dullo:2020} predict $\mathcal{M}_\bullet=4.89\pm0.85$, with $P(\mathcal{M}_\bullet\leq5)=55.1\%$, and $\mathcal{M}_\bullet=4.59\pm0.85$, with $P(\mathcal{M}_\bullet\leq5)=68.5\%$, from $\mathcal{C}_\mathrm{FUV,tot}$ and $\mathcal{C}_\mathrm{NUV,tot}$, respectively.
Also, we can apply the $M_\bullet$--$M_{\rm gal,\star}$ relation \citep[][equation~3]{Davis:2018} to the total stellar mass of UGC~8153 ($\mathcal{M}_{\rm gal,\star}=9.78\pm0.16$) and predict a black hole mass of $\mathcal{M}_\bullet=4.11\pm0.98$, with $P(\mathcal{M}_\bullet\leq5)=81.7\%$.
Thus, the color-based and total stellar mass predictions for UGC~8153 are consistent with our predictions for an IMBH, $\mathcal{M}_\bullet=4.13\pm0.55$, with $P(\mathcal{M}_\bullet\leq5)=94.4\%$.

Similarly for \textbf{UGC~10020}, \citet{Dullo:2020} predict $\mathcal{M}_\bullet=5.46\pm0.85$, with $P(\mathcal{M}_\bullet\leq5)=29.4\%$, and $\mathcal{M}_\bullet=5.35\pm0.85$, with $P(\mathcal{M}_\bullet\leq5)=34.1\%$, from $\mathcal{C}_\mathrm{FUV,tot}$ and $\mathcal{C}_\mathrm{NUV,tot}$, respectively.
Furthermore, we can apply the $M_\bullet$--$M_{\rm gal,\star}$ relation \citep[][equation~3]{Davis:2018} to the total stellar mass of UGC~10020 ($\mathcal{M}_{\rm gal,\star}=9.59\pm0.22$) and predict a black hole mass of $\mathcal{M}_\bullet=3.55\pm1.14$, with $P(\mathcal{M}_\bullet\leq5)=89.9\%$.
Likewise, the color-based total stellar mass predictions for UGC~10020 are consistent with our predictions for an IMBH, $\mathcal{M}_\bullet=4.98\pm0.47$.
However, the higher color-based mass predictions for UGC~10020 are in line with our weak $P(\mathcal{M}_\bullet\leq5)=52.0\%$, which is the least certain IMBH candidate among our sample of 23 targets of interest.

\section{Discussion}\label{sec:discussion}

\subsection{The Prototypical Sd Galaxy}\label{sec:ave}

We observe that the geometry of the spiral arm shape in Sd galaxies is not always as loosely wound as one might expect.
Some of this is due to obvious misclassifications of the morphological type.\footnote{Consulting the meta-analysis of morphological types for our galaxies from HyperLeda, we find that only two of our galaxies have morphological types that do not agree ($T\neq7$) with our adopted classifications from the \citetalias{RC3}: UGC~283, $T=5.8\pm0.7$ and UGC~10146, $T=8.7\pm1.4$.}
For example, our smallest pitch angle measurement is $|\phi|=5\fdg8\pm2\fdg8$ for UGC~1702, which is an absurdly small pitch angle for a legitimate Sd galaxy.
However, UGC~1702 is also one of the most massive galaxies in our sample, with $\mathcal{M}_{\rm gal,\star}=10.55\pm0.21$.
Moreover, it has a high rotational velocity with $v_\mathrm{max}=283.8\pm14.3$\,km\,s$^{-1}$.
Indeed, our trivariate relation predicts the most massive SMBH in our sample, with a mass of $\mathcal{M}_\bullet=8.61\pm0.37$, with $P(\mathcal{M}_\bullet\leq5)\approx0\%$.

In some instances, the geometry of arms may not agree well with a weak bulge.
The \citetalias{RC3} bases their classifications on the \citet{Vaucouleurs:1959} classification approach, which is based the appearance of the spiral arms and the bulge of a galaxy.
They consider both (\emph{i}) the degree of openness (\textit{i.e.}, $|\phi|$) and (\emph{ii}) the resolution of spiral arms into star clusters or very luminous stars (\textit{i.e.}, knotty vs.\ smooth spiral structure), and additionally (\emph{iii}) the relative prominence of the bulge or central concentration (\textit{i.e.}, $B/T$).\footnote{
\citet{Willett:2013} and \citet{Masters:2019} argue that modern morphological classifications of spiral galaxies have devolved the traditional tenets of the classic Hubble-Jeans sequence \citep{Jeans:1919,Jeans:1928,Lundmark:1925,Hubble:1926b,Hubble:1926,Hubble:1927,Hubble:1936} that prioritized spiral arm winding \citep{Bergh:1998,Lapparent:2011}, but still considered bulge size; contemporary morphological sequences are now predominantly ordered on central bulge size alone, with no reference to spiral arms \citep{Graham:2008,Willett:2013}.}
These criteria that govern the stage for spirals may be inconsistent in some cases or may be overruled by other factors that affect the morphological type \citep{Sandage:1961,Sandage:1994}.
Importantly for our study of Sd types, \citet{Mengistu:2023} find that bluer and lower mass galaxies most closely follow the ``expected'' arm windiness correlation with bulge size, \textit{i.e.}, smaller bulges with loosely wound spiral arms.

The $B/T$ ratio is directly related to the morphological type, in general, but with considerable scatter for a given type \citep{Simien:1986,Laurikainen:2007,Graham:2008,Willett:2013}.
Specifically, \citet{Masters:2019} find that galaxies with larger bulges favor tighter spiral arms, while those with smaller bulges have a wide range of arm winding; \textit{cf.}\ \citet{Lingard:2021}, who find no correlation between bulge size and pitch angle.
Very recently, \citet{Chugunov:2023} affirmed ``that the pitch angle of spiral arms decreases with increasing bulge or bar fraction.''
\citet{Smith:2022} found that blue (\textit{i.e.}, star-forming) galaxies predominantly exhibit loosely wound spiral arms and red (\textit{i.e.}, quiescent) galaxies mainly display tightly wound spiral arms.
Similarly, late-type spirals have been shown to have stronger arms \citep{Yu:2020} and higher star-formation rates \citep{Yu:2021}.\footnote{\citet{Aktar:2023} found no evidence of a trend between star-formation rates and pitch angle, but postulated that the lack of a correlation ``may be explained by different star formation efficiencies caused by the distinct galactic ambient conditions.''}

Importantly, multiple studies find a general trend (albeit with considerable scatter) between pitch angle and Hubble stage, \textit{i.e.}, $|\phi|\propto T$ \citep{Kennicutt:1981,Seigar:1998,Ma:1999,Baillard:2011,Yu:2018,Diaz-Garcia:2019,Yu:2019,Yu:2020,Chugunov:2023}.
More precisely, \citet{Treuthardt:2012} show that the correlation between $|\phi|$ and $T$ is tightest when selecting spiral galaxies with fast rotating bars, which is in close agreement with the theoretical relation \citep{Roberts:1975}.
Notably, \citet{Yu:2019} find that $|\phi|$ is most closely correlated with $M_{\rm gal,\star}$, especially so for low-mass galaxies.
Overall, there are related connections between multiple parameters, with positive correlations between $\phi$--$T$ and $\phi$--(absolute magnitude) relations \citep{Kennicutt:1981}, and negative correlations between $\phi$--$v_\mathrm{max}$ \citep{Kennicutt:1981,Davis:2019} and $v_\mathrm{max}$--$T$ \citep{Roberts:1978} relations.

Frustratingly, if the image resolution is inadequate, the degree of openness cannot be accurately determined, and the default is toward a later type.
Indeed, \citet{Peng:2018} empirically found that $|\phi|\propto z$, that is, the average pitch angle observed in galaxies is perceived to increase (loosen) as a function of redshift.
Although, this is not expected to be an intrinsic effect, but rather a systematic effect of ``tightly wound arms becoming less visible as image quality degrades'' \citep{Peng:2018}.
Hence, some of the more distant galaxies in our sample of ``Sd'' galaxies could have tightly wound spiral arms that were not accounted for in classification catalogs from decades ago due to lower image resolutions.
From that point of view, the ($T=7.0$) morphological types for our sample could be considered upper limits.
Additionally, \citet{Graham:2008} point out that disk luminosities become progressively dimmer with increasing Hubble type, further exacerbating efforts to resolve the geometry of spiral arms.
As telescope technology continues to advance, it becomes increasingly probable that new, high-resolution surveys may overturn old morphological classifications if they can better resolve the geometry of the spiral arms.

Even though the Sd sample tends towards large pitch angles, the broad range they cover could be a possible indication of their diverse origins or evolutionary histories.
The gas content in Sd galaxies\footnote{\citet{Lacerda:2020} found that Sd galaxies have some of the highest gas fractions ($f_{\rm gas}$) of any morphological type (see their figure~9 and table~3), with an average $f_{\rm gas}=6.76\%$, which is almost three times the average proportion they found in their Sa galaxies ($f_{\rm gas}=2.40\%$).} could also have a role to play in diminished pitch angle values.
\citet{Davis:2015} observed that, due to spiral density wave theory \citep[\textit{i.e.},][]{Lin:1966}, the pitch angle of spiral galaxies is due to both the central mass of a galaxy (which includes the central black hole), as well as the density of the disk of a galaxy, such that the pitch angle is directly proportional to the disk density and inversely proportional to the central mass.
\citet{Graham:2008} show that the surface density of disks decrease as the morphological type of a galaxy becomes increasingly later.
This can become problematic in accounting for later morphological types in the galaxy stellar mass function; \citet{Kim:2022} estimate that the majority of low-surface brightness galaxies are missed in redshift surveys at $z>0.9$.
Indeed, the problem is not easily ignorable since the number density of galaxies \emph{increases} monotonically as their stellar masses \emph{decrease} \citep{Driver:2022}.
Moreover, the prominence of low stellar-mass galaxies becomes increasingly relevant as the slope of the galaxy stellar mass function is shown to steepen with redshift \citep{Navarro-Carrera:2023}.

Thus, if the ratio of disk density to central mass does not decrease at the same rate as the morphological type increases, the pitch angle of an Sd galaxy could very well be lower than expected if it were only due to the central mass.
Another complication to consider is the environment of each galaxy.
Cluster galaxies can experience ram-pressure stripping, which ``unwinds'' their spiral arms \citep{Bellhouse:2021}.
Although, this effect would act to loosen (\textit{i.e.}, increase $|\phi|$) the pitch angle of a galaxy.
Therefore, our pitch angle measurements could be skewing the predicted black hole masses towards higher masses in some cases, depending on the influence of unknown densities.

We are also cognizant of the possibility that other spiral genesis mechanisms besides the spiral density wave theory could be at play, \textit{e.g.}, swing amplification \citep[see][for a review]{Dobbs:2014}.
Notably, \citet{Hart:2018} found that $\approx$40\% of the galaxies in their sample have spiral arms that can be modeled by swing amplification.
\citet{Yu:2020} compared their work with the models of \citet{Hart:2018} and speculated that later morphologically-typed spiral galaxies in their sample could be similarly influenced by swing amplification.
Following work by \citet{Yu:2018b} and \citet{Pringle:2019}, \citet{Lingard:2021} found that their sample of spiral galaxies could be explained by evolution of transient/recurrent spirals via swing amplification that wind up over time, \emph{if} pitch angles are sufficiently high.
Similarly, \citet{Reshetnikov:2023} showed from observations that $|\phi|\propto z$, \textit{i.e.}, pitch angles tend to decrease (windup) with time. 
Notably, \citet{Hart:2017} found that central mass concentration alone does not govern pitch angle; they found that galaxies which are more disk dominated contain more spiral arms with tighter pitch angles.

It is also notable that there appears to be an empirical correlation between pitch angle and dark matter halos; \citet{Seigar:2005,Seigar:2006,Seigar:2014} have demonstrated an anticorrelation between pitch angle and the central mass concentration of a spiral galaxy via measurement of the rate of shear of its rotation curve ($\Gamma$).\footnote{We note that subsequent studies \citep{Kendall:2015,Yu:2018,Yu:2019} have also found an anticorrelation between $|\phi|$ and $\Gamma$, albeit a much weaker anticorrelation with higher scatter.}
Additionally, simulations have shown that late-type bulgeless galaxies pose an enigma due to an apparent dichotomy between their observed and simulated angular momenta.
\citet{D'Onghia:2004} demonstrated that in the absence of major mergers, dark matter halos have too low an angular momentum to reconcile the observed disks of their embedded bulgeless late-type galaxies.
Indeed, \citet{Rodriguez-Gomez:2022} found that galaxies with higher specific angular momenta reside in faster spinning halos and tend to host less massive black holes.
Also, \citet{Rodriguez-Gomez:2022} further described that halo spin is anti-correlated with black hole mass at fixed galaxy or halo mass.
Such complications could be an additional factor in discrepancies between our observed properties (\textit{i.e.}, pitch angle, stellar mass, and rotational velocities).
This further warrants and justifies our adoption of a higher-dimensional black hole mass predictor, rather than relying on only one two-dimensional scaling relation parameterization \citep[see also][]{Williams:2023}.

Overall, we find that the notion of a truly average Sd galaxy is false.
Similarly, \citet{Daniels:1952} found, in his landmark anthropometric study, that the \emph{average man} does not exist.
From a study of 4,063 men, \citet{Daniels:1952} found that no man was average (\textit{i.e.}, within $\pm0.3$ standard deviations of the mean) across more than nine out of the 132 body measurements performed for the study.
Using the same definition of average, we find that only three of our 85 galaxies are ``average'' across $|\phi|$, $v_\mathrm{max}$, and $\mathcal{M}_{\rm gal,\star}$.
These most average galaxies include: UGC~384, 2109, and 3074.
Together, these three prototypical Sd galaxies have a modal $\mathcal{M}_\bullet=5.84\pm0.19$, with $P(\mathcal{M}_\bullet\leq5)=0.3\%$.
Therefore, IMBHs are likely to be harbored only in Sd galaxies that are significantly looser wound, rotating slower, and/or less massive than the exemplar of Sd galaxies.

\subsection{Sd Classifications: Optical vs.\ Mid-IR}

In order to draw comparisons of our population of Sd galaxies, we sought an independent study that had a statistically viable number of Sd galaxies with quantitative pitch angle measurements.
The largest such sample we could find comes from the \textit{Spitzer} Survey of Stellar Structure in Galaxies \citep[][hereafter \citetalias{Sheth:2010}]{Sheth:2010}.
A series of papers \citep{Buta:2010,Buta:2015} goes through careful morphological classifications of their 2,352 galaxies.
Using the classifications of \citet{Buta:2015} and pitch angle measurements by \citet{Herrera-Endoqui:2015}, refined by \citet{Diaz-Garcia:2019}, we identified 22 Sd galaxies with pitch angle measurements from the \citetalias{Sheth:2010} sample.

\begin{figure}
\includegraphics[clip=true, trim= 0mm 0mm 0mm 0mm, width=\columnwidth]{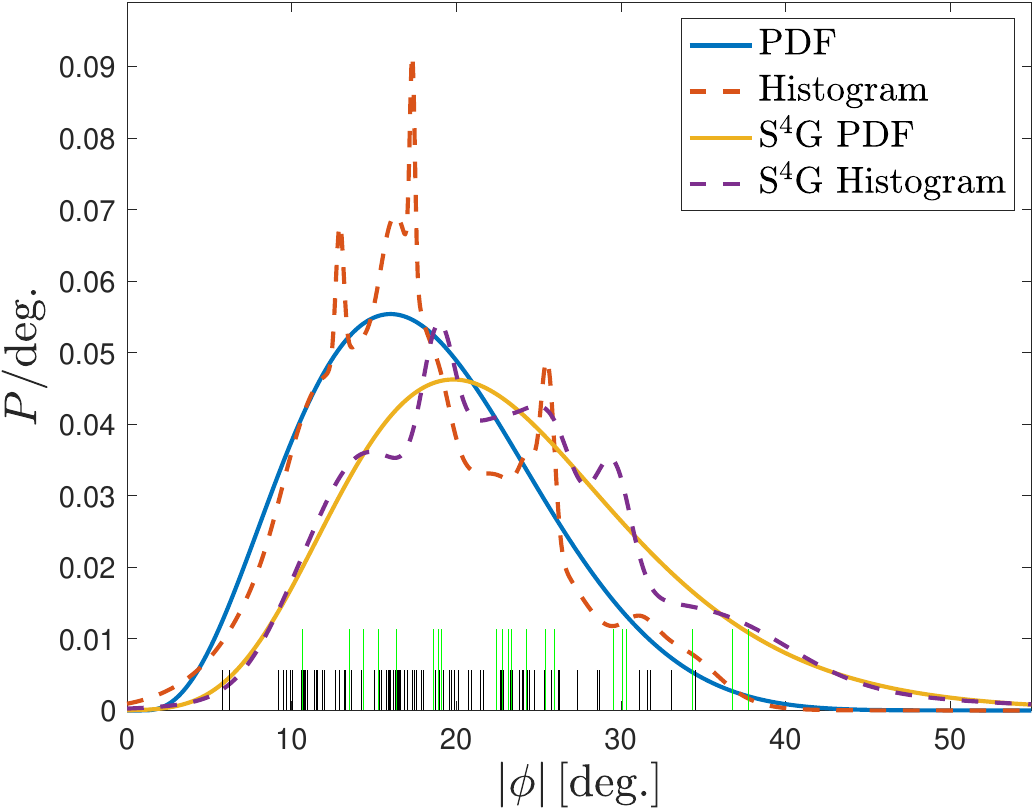}
\caption{The distribution of the pitch angles ($|\phi|$) for our sample of 85 Sd galaxies (from Table~\ref{tab:sample}, Column~6) is shown as a rug plot with short \textbf{black} tassels with a mean $|\phi|$ of $18\fdg0\pm7\fdg0$ and a median of $17\fdg1\pm4\fdg7$.
The smoothed histogram (\textcolor{MATLAB_red}{{\hdashrule[0.35ex]{8mm}{1pt}{1mm}}}) is generated from the summation of all 85 pitch angle measurements.
The fitted PDF (\textcolor{MATLAB_blue}{{\hdashrule[0.35ex]{8mm}{0.4mm}{}}}) peaks at $|\phi|=16\fdg0\pm0\fdg8$.
We compare this with a sample of 22 Sd galaxies (\textcolor{MATLAB_purple}{{\hdashrule[0.35ex]{8mm}{1pt}{1mm}}}) shown as a rug plot with long \textcolor{green}{green} tassels with pitch angle measurements from the \citetalias{Sheth:2010} sample \citep{Herrera-Endoqui:2015,Diaz-Garcia:2019}, with a mean $|\phi|$ of $23\fdg3\pm9\fdg4$, a median of $22\fdg2\pm6\fdg0$, and a fitted PDF (\textcolor{MATLAB_orange}{{\hdashrule[0.35ex]{8mm}{0.4mm}{}}}) that peaks at $|\phi|=19\fdg8\pm2\fdg0$.
}
\label{fig:pitch}
\end{figure}

In Figure~\ref{fig:pitch}, we present a comparison of the pitch angle distribution of our 85 Sd galaxies alongside the 22 Sd galaxies from \citet{Diaz-Garcia:2019}.
Both distributions demonstrate a similar shape (\textit{i.e.}, positive skewness), but the peaks are notably different: $|\phi|=16\fdg0\pm0\fdg8$ for our galaxies and $|\phi|=19\fdg8\pm2\fdg0$ for the galaxies from the \citetalias{Sheth:2010} sample.
We performed a Kolmogorov--Smirnov (K--S) test \citep{Kolmogorov:1933,Smirnov:1948} to assess the likelihood that both samples come from the same distribution.
In doing so, we found a $p$-value $=0.0110$, thus rejecting the null hypothesis at the 1.10\% level (\textit{i.e.}, a 98.9\% probability that the samples come from different parent populations).

However, this is not the end of the story for this comparison.
Crucially, our 85 galaxies were classified as Sd galaxies by the \citetalias{RC3} based on $B$-band images, whereas the 22 galaxies classified as Sd galaxies by \citet{Buta:2015} were based on middle-infrared (mid-IR) images.
Specifically, \citet{Buta:2015} exclusively used 3.6-$\micron$ images, which highlight the photospheric light of old stars \citep{Pahre:2004}.
This can lead to a complicated difference in the perceived morphologies that were traditionally observed in $B$-band light that highlights young stellar populations and is strongly affected by extinction and reddening.
Moreover, \citet{Buta:2015} also used an updated morphological notation system \citep[\textit{e.g.},][]{Buta:2014} that is similar to, but more extensive than, the notational system provided in the \citetalias{RC3}.

Overall, it is generally accepted that the morphological classification schemes derived from $B$-band observations can be effectively applied to infrared images and reproduce the gamut of morphological diversity \citep{Eskridge:2002,Buta:2010,Buta:2015}.
However, it has been claimed that optical and near-IR morphologies of spiral galaxies are uncorrelated; \citet{Block:1999} found that ``the Hubble tuning fork does not constrain the morphology of the old stellar Population~II disks.''
Moreover, \citet{Block:1999} found extreme examples where galaxies on opposite ends of the Hubble tuning fork in optical light can have the same morphology when observed in the near-infrared ($K^\prime$-band).
Nonetheless, \citet{Eskridge:2002} found that $B$-band vs.\ $H$-band morphologies are mostly similar.
Separately, \citet{Buta:2010,Buta:2015} found that $B$-band vs.\ 3.6-$\micron$ morphologies are also largely the same.
Although, systematic differences are observed.\footnote{For a further investigation of how pitch angle measurements vary as a function of the wavelength of light and comparisons with predictions of the density wave theory, see the following papers \citep[\textit{e.g.},][]{Pour-Imani:2016,Yu:2018,Miller:2019,Abdeen:2020,Abdeen:2022,Garcia:2023,Chen:2023c}.
}

Infrared morphologies exhibit a clear ``earlier effect'' \citep{Eskridge:2002,Buta:2010,Buta:2015}, which is a result of the increased prominence of the bulge and the decreased prominence of star-forming regions in spiral arms when observed in the infrared.
Indeed, the extensive multi-wavelength bulge-disk decomposition study of \object{M81} (NGC~3031) by \citet{Gong:2023} showed that the S\'ersic index \citep{Sersic:1968} and effective radius of its bulge are proportional to wavelength, so much so that M81 appears to have a prominently classical bulge in the infrared and bulgeless at ultraviolet wavelengths.
However, \citet{Ito:2023} saw a negative correlation between the observed wavelength and effective radius of $z\geq3$ quiescent galaxies, and their resulting size--mass relation is lower than those observed at lower redshifts \citep[\textit{e.g.},][]{Wel:2014,Hon:2022,Hon:2023}.
See also \citet{Yao:2023}, \citet{Ono:2023}, and \citet{Ormerod:2023}, for their studies of the variation of morphological parameters with rest-frame wavelength.
Thus, because the bulges visually stand out more in IR images, they tend to be classified as an earlier type than they were in $B$-band images.
Specifically, \citet{Eskridge:2002} find that their $H$-band classifications are systematically one stage earlier than the \citetalias{RC3} $B$-band classifications on average ($T_H=T_B-1$).
Therefore, it is plausible that the 22 \citetalias{Sheth:2010} $T_{3.6\,\micron}=7$ galaxies that we compare with in Figure~\ref{fig:pitch} are actually $T_B=8$ galaxies with expectedly higher pitch angles, whose classifications are being overruled by their more prominent bulges when observed in the mid-IR.

We identified one of the 22 \citetalias{Sheth:2010} galaxies (NGC~5668) also in our sample that was independently measured by \citet{Herrera-Endoqui:2015} and \citet{Diaz-Garcia:2019}.
\citet{Diaz-Garcia:2019} measured a pitch angle of $|\phi|=29\fdg7\pm3\fdg9$, which is slightly higher, but consistent with our determination of $|\phi|=27\fdg3\pm2\fdg8$ for NGC~5668.
However, \citet{Buta:2015} classify NGC~5668 as SAB(rs)cd, or $T_{3.6\,\micron}=6.5$.
Thus, supporting our suspicions of the earlier effect in its classification, although the pitch angles remain consistent in this case.
When ultimately analyzed in aggregate, the statistically-significant observation of larger absolute pitch angles for the 3.6-$\micron$ sample confirms the implications of our aforementioned K--S test that the samples are drawn from different populations.
Verily, comparison of morphologies across the electromagnetic spectrum is problematic and confounds demographical comparisons of galaxies.

\subsection{Comparison with a General Spiral Population}

In their study of a volume- and magnitude-limited sample of 140 spiral galaxies, \citet{Davis:2014} found that the black hole mass function (BHMF) of spiral galaxies peaks at $1.17\times10^7\,\mathrm{M}_\sun$.
Their volume-limited sample of 140 spiral galaxies is comprised of Sa, Sab, Sb, Sbc, Sc, Scd, Sd, and Sm types.
However, it included only one ($\approx$1\%) Sd galaxy (\object[ESO138-010]{ESO~138-010}); the most common morphological type was Sc (34\%; see their figure~3).
For comparison with their BHMF, we have produced a distribution of our black hole mass estimates (Figure~\ref{fig:all}).\footnote{
In this and subsequent figures, the histograms are the summation of individual normal distributions that describe each measurement.}
From Figure~\ref{fig:all}, we find that the black hole mass distribution of Sd galaxies is indeed significantly different from the general spiral galaxy sample of \citet[][their figure~7]{Davis:2014}.
Our fitted PDF peaks at $1.01\times10^6\,\mathrm{M}_\sun$, or $8.6\%$ of the most probable black hole mass found in an average (Sc) spiral galaxy.

\begin{figure}
\includegraphics[clip=true, trim= 0mm 0mm 0mm 0mm, width=\columnwidth]{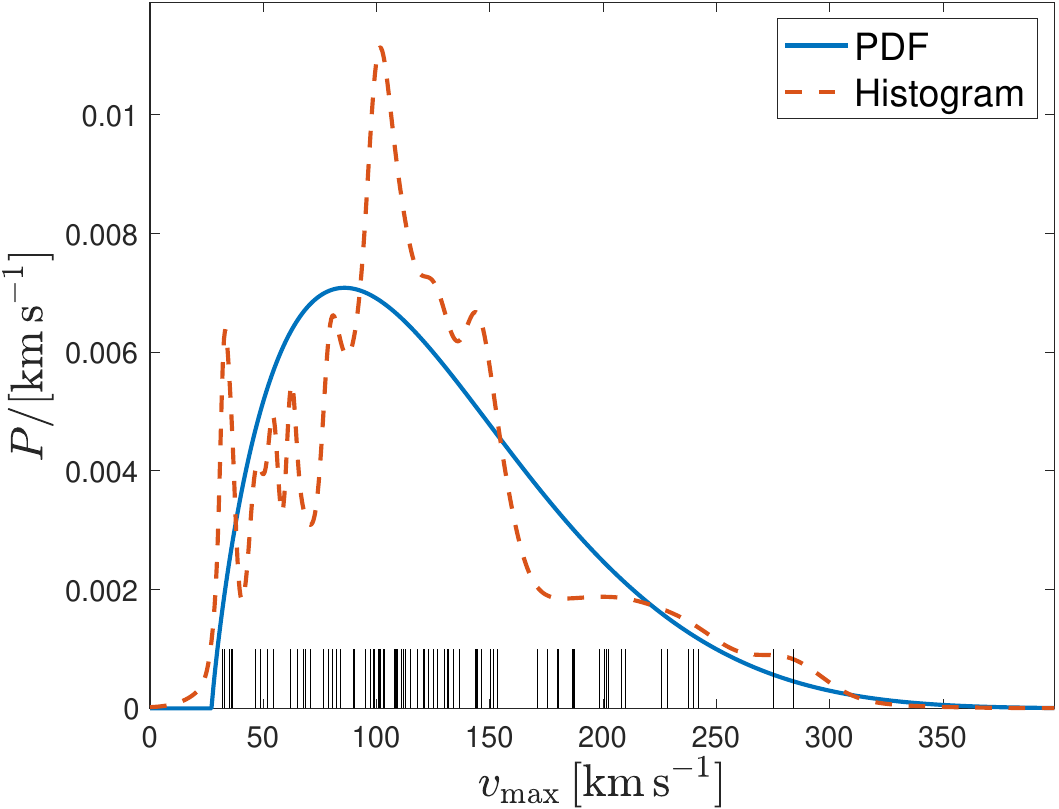}
\caption{The distribution of the maximum rotational velocities ($v_\mathrm{max}$) for all 85 Sd galaxies (from Table~\ref{tab:sample}, Column~10) is shown as a rug plot along the bottom axis with a mean $v_\mathrm{max}$ of $124.9\pm60.2$\,km\,s$^{-1}$ and a median of $114.3\pm32.9$\,km\,s$^{-1}$.
The smoothed histogram (\textcolor{MATLAB_red}{{\hdashrule[0.35ex]{8mm}{1pt}{1mm}}}) is generated from the summation of all 85 $v_\mathrm{max}$ measurements.
The fitted PDF (\textcolor{MATLAB_blue}{{\hdashrule[0.35ex]{8mm}{0.4mm}{}}}) peaks at $v_\mathrm{max}=85.8\pm6.5$\,km\,s$^{-1}$.}
\label{fig:vrot}
\end{figure}

\begin{figure}
\includegraphics[clip=true, trim= 0mm 0mm 0mm 0mm, width=\columnwidth]{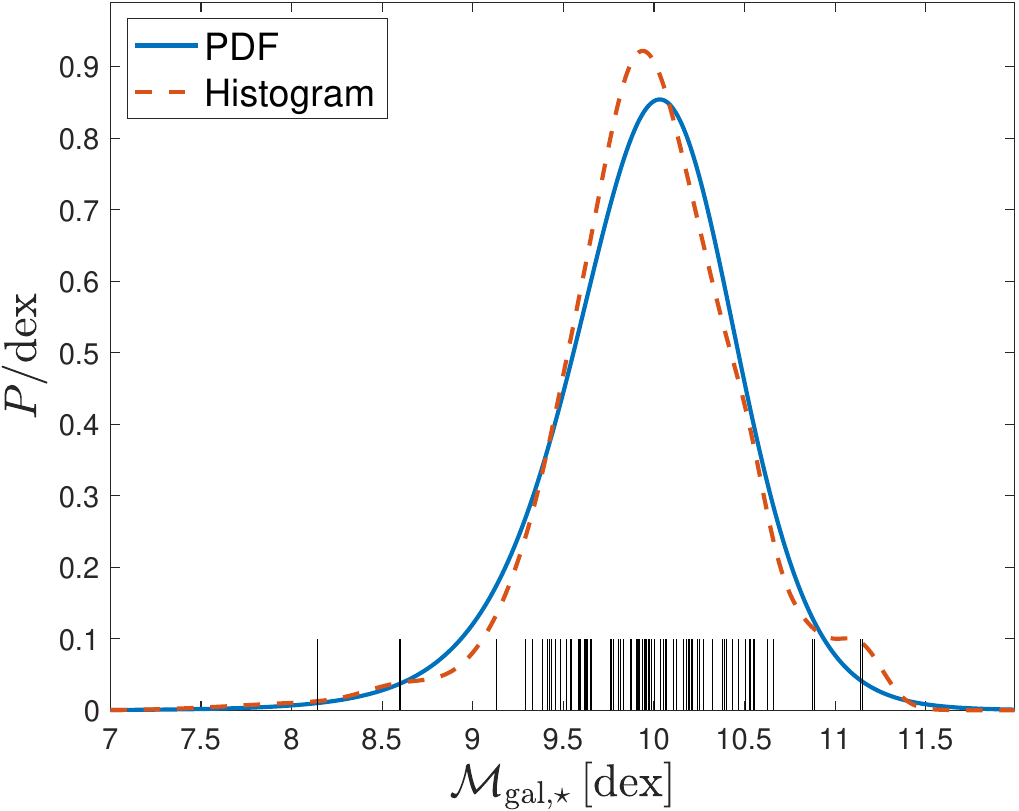}
\caption{The distribution of the total galaxy stellar masses ($\mathcal{M}_{\rm gal,\star}$) for all 85 Sd galaxies (from Table~\ref{tab:sample}, Column~8) is shown as a rug plot along the bottom axis with a mean $\mathcal{M}_{\rm gal,\star}$ of $9.96\pm0.53$ and a median of $9.97\pm0.30$.
For comparison, \citet{Lacerda:2020} found an average value of $\mathcal{M}_{\rm gal,\star}=9.90\pm0.09$ from their sample of 34 Sd galaxies.
The smoothed histogram (\textcolor{MATLAB_red}{{\hdashrule[0.35ex]{8mm}{1pt}{1mm}}}) is generated from the summation of all 85 $\mathcal{M}_{\rm gal,\star}$ measurements.
The fitted PDF (\textcolor{MATLAB_blue}{{\hdashrule[0.35ex]{8mm}{0.4mm}{}}}) peaks at $\mathcal{M}_{\rm gal,\star}=10.03\pm0.06$.}
\label{fig:Mgal}
\end{figure}

We also present the distributions of pitch angles (Figure~\ref{fig:pitch}), rotational velocities (Figure~\ref{fig:vrot}), and total stellar masses (Figure~\ref{fig:Mgal}).
As we can see, these distributions exhibit slight telltale evidence of multiple populations also reflected by the multimodal distribution of predicted black hole masses in Figure~\ref{fig:all}.
These similarities across independent measurements give credence to the seemingly disparate spiral geometries uncovered by our pitch angle measurements across this sample of only Sd galaxies.\footnote{
See also the recent meta-analysis performed by \citet{Savchenko:2020}, whose figure~13 demonstrates significant variance of pitch angle across morphological types.}
We note that our general PDF fit to the pitch angle distribution in Figure~\ref{fig:pitch} is not appreciably different from the general shape of the distribution for all spiral types \citep[][figure~6]{Davis:2014}.\footnote{The modes of both the distributions for Sd ($|\phi|=16\fdg0$) and all ($|\phi|=18\fdg5$) spiral galaxies are remarkably close to the pitch angle ($|\phi|\approx17\fdg0$) of the golden spiral \citep[][appendix~A]{Davis:2014}.}
However, the histogram for Sd galaxies in Figure~\ref{fig:pitch} does exhibit a subpopulation of high pitch angle galaxies (\textit{e.g.}, the prominence at $|\phi|=25\fdg5$), and lacks the enhanced population of galaxies with $|\phi|\lesssim10\degr$ found in \citet{Davis:2014}.
This result is also reflected in the recent work of \citet{Fusco:2022}, who conducted a follow-up study to \citet{Davis:2014} by analyzing the complementary population of 74 low-mass galaxies (peak probability at $|\phi|=17\fdg5$) that was excluded from \citet{Davis:2014}.
\citet{Fusco:2022} report a similar enhancement to the BHMF from Scd--Sm galaxies, and show that galaxies in this morphological population are predominantly the hosts of ``less-than-supermassive'' black holes ($M_\bullet\lesssim10^6\,\mathrm{M_\sun}$).

\subsection{Implications}

\subsubsection{Seeding Models}

Astrophysicists are grappling with the puzzle of how SMBHs formed within the first billion years of the Universe's existence (at $z\gtrsim6$).
These SMBHs, some exceeding a billion solar masses \citep{Tripodi:2023}, challenge conventional models of black hole formation from stellar collapse.
Observations of quasars at high redshifts support the existence of these massive black holes \citep{Fan:2001,Fan:2003,Willott:2007,Mortlock:2011,Morganson:2012,Kashikawa:2015,Wu:2015,Banados:2016,Banados:2018,Decarli:2018,Matsuoka:2019,Onoue:2019,Endsley:2023}.
The prevailing theory suggests that ancient ``seeds'' of black holes underwent rapid growth through accretion and mergers \citep[\textit{e.g.},][]{Soltan:1982,Small:1992,Kauffmann:2000,Volonteri:2005,Shankar:2009,Shen:2020,Lin:2022,Li:2023}.
Theorists employ simulations and analytic studies to explore the possible scenarios leading to the formation of these massive black holes \citep{Dijkstra:2014,Visbal:2014,Habouzit:2016a,Habouzit:2016b,Lupi:2021,Trinca:2022,Bhowmick:2022,Bennett:2023,Jeon:2023}, albeit with varying success rates \citep[see][for a review]{Matteo:2023}.

While the conventional pathway for SMBH formation involves the collapse of massive stars, episodic periods of super-Eddington accretion \citep[\textit{e.g.},][]{Madau:2014,Jiang:2019,Massonneau:2022}, as well as exotic theories involving primordial black holes, inflation, and dark matter have been proposed \citep[\textit{e.g.},][]{Clesse:2015,Chen:2022,Davoudiasl:2022,Hooper:2023}.
Evidence from element abundance ratios and radiation from metal-poor galaxies indicates the existence of supermassive stars in the early universe \citep{Kojima:2021}, which could have collapsed into IMBHs \citep{Ohkubo:2006}.
Various mechanisms, such as direct collapse \citep[\textit{e.g.},][]{Loeb:1994,Bromm:2003,Begelman:2006,Lodato:2006,Hosokawa:2013,Ferrara:2014,Umeda:2016,Haemmerle:2018a,Haemmerle:2018b,Latif:2020,Moriya:2021}, collisional mergers in dense clusters \citep{Portegies_Zwart:2002,Davies:2011,Lupi:2014,Das:2021,Rose:2022,Vergara:2022} or successive black hole mergers \citep{Fragione:2021,Fragione:2022}, and dynamical interactions in young stellar clusters \citep{Carlo:2021,Torniamenti:2022}, are proposed for the formation of IMBH seeds.

Observational efforts, including the \textit{James Webb} Space Telescope \citep[JWST;][]{Gardner:2006}, aim to uncover evidence supporting or refuting these black hole seeding models.
The detection and study of IMBHs will play a crucial role in distinguishing between different evolutionary pathways and understanding the origins of the first SMBH seeds \citep{Greene:2012,Ricarte:2018,Sassano:2021,Regan:2022}.
Advances in observational technologies promise to shed more light on this enigmatic aspect of astrophysics in the coming decades.

\subsubsection{Gravitational Radiation and Lenses}

Recent advancements in gravitational-wave astronomy, particularly through instruments like the Laser Interferometer Gravitational-wave Observatory \citep[LIGO;][]{Abramovici:1992,Abbott:2009}, have opened new avenues for detecting IMBHs.
LIGO's advanced configuration \citep{Harry:2010,aLIGO:2015} allows it to potentially observe the formation of IMBHs from runaway merger processes in dense NSCs \citep{Fragione:2020,Fragione:2020b,Fragione:2020c,Escala:2021} or mergers involving black holes from Population~III stars \citep{Fryer:2001,Madau:2001,Haiman:2001,Heger:2003,Volonteri:2003,Abadie:2012,Abbott:2017,Kovetz:2018,Antonini:2019,Wang:2022}.
While earlier observing runs yielded no IMBH merger candidates \citep{Abbott:2017b,Abbott:2019,Vajpeyi:2022}, recent years have seen the detection of potential IMBHs, with masses around $156.3^{+36.8}{-22.4}$\,M$_\sun$ \citep[GW190521;][]{Abbott:2021} and possibly $172.9^{+37.7}{-33.6}$\,M$_\sun$ \citep[][\textit{cf.}\ \citealt{Abbott:2022}]{LIGO:2021}.
These detections, such as the gravitational-wave event GW190521, prompt speculation about their origins, including mergers within ultradwarf galaxies \citep{Palmese:2021} or intermediate-mass ratio inspirals involving IMBHs \citep{Fishbach:2020,Nitz:2021}.

Future ground-based observatories like the Cosmic Explorer \citep{Evans:2021} and space-based missions like the Laser Interferometer Space Antenna \citep[LISA;][]{LISA:2017,LISA:2022} hold promise for further exploring the population of IMBHs across different masses and redshifts.
Particularly, modeling by \citet{Izquierdo-Villalba:2023} shows that the LISA will detect coalescing massive black hole binaries with total masses of 10$^4$--10$^5$\,M$_\sun$ in host galaxies that are predominantly ``extreme late-types'' (\textit{i.e.}, akin to our Sd galaxies) at $z=1.5$--3.0.\footnote{For this endeavor, the \textit{Roman} Space Telescope will be able to aid in finding precursors, \textit{i.e.}, massive black hole binaries before their separations are quite small enough to be detected via gravitational waves with the LISA \citep{Haiman:2023}.}
Additionally, collaborations between observatories, such as joint observations between LISA and TianQin, will enhance our understanding of these enigmatic objects \citep{Alejandro:2023}.
Furthermore, combining gravitational-wave observations with electromagnetic data is crucial for a comprehensive understanding of IMBHs, including their distribution and potential formation mechanisms \citep{Saini:2022,Piro:2023,Bardati:2023}.
Microlensing \citep{Lam:2022,Sahu:2022} and studies of globular clusters \citep{Kains:2016} offer complementary methods for detecting and studying IMBHs, shedding light on their prevalence and dispersal throughout the Universe.

Emerging evidence indicates that the true population is significant and that the Universe might be rife with IMBHs.
Recently, \citet{Paynter:2021} presented a novel way to search for and find IMBHs as the intervening gravitational lenses to distant gamma ray bursts (GRBs).
They claim to have discovered an IMBH at $z\approx1$, which is lensing a background GRB at $z\approx2$.
Depending on the unknown redshift, they estimate the mass of the lensing object to be $\approx$$5.5\times10^4\,\mathrm{M_\sun}$, which they interpret as evidence for an IMBH.
Based upon the observed frequency of lensed GRBs, they estimate that the present-day number density of IMBHs with masses between $\approx$$10^4$--$10^5\,\mathrm{M_\sun}$ is $\approx$$2.3\times10^3\,\mathrm{Mpc}^{-3}$.
Compared with our aforementioned estimate of $>$$4.96\times10^{-6}$\,Mpc$^{-3}$ for central IMBHs, this implies that the \emph{vast} majority of IMBHs are not in the center of galaxies.
Overall, these multi-faceted approaches contribute to unraveling the mysteries surrounding IMBHs and their role in shaping the cosmos.

\subsubsection{Star Clusters}

Various astronomical phenomena, including young, globular, and nuclear star clusters, contribute significantly to black hole formation \citep{Mapelli:2021,Mapelli:2021b}.
NSCs play crucial roles in the hierarchical mergers of black holes, resulting in distinct merger rates and masses \citep[see][for a review of NSCs]{Neumayer:2020}.
Young dense massive star clusters rapidly produce IMBHs within 15 million years \citep{Rizzuto:2021}, covering a wide mass range observed in gravitational-wave detections \citep{Rizzuto:2022}.
Gas-rich NSCs act as long-lived channels continuously forming IMBHs throughout cosmic time \citep{Natarajan:2021}.

Simulations show that nucleated dwarf galaxies are prime environments for efficient IMBH mergers \citep{Khan:2021}.
Tidal stripping of nucleated dwarf galaxies yield ultracompact dwarf galaxies \citep[\textit{e.g.},][]{Ferrarese:2016,Graham:2019c,Pechetti:2022,Dumont:2022}, which can ultimately become the captured nuclei of previously bulgeless galaxies via mergers, with their IMBHs in tow \citep{Graham:2021b}.
X-ray activity in the centers of low-mass galaxies, like those in our sample, can indicate the presence of nuclear IMBHs \citep{Graham:2019,Graham:2021}.

Most galactic nuclei harbor SMBHs, with a significant fraction coexisting with NSCs \citep[\textit{e.g.},][]{Sruthi:2023}.
However, not all nucleated galaxies possess SMBHs.
\citet{Askar:2021,Askar:2022} explore factors influencing the occupation fraction of SMBHs in NSCs .
Gravitational-wave recoil kicks can affect SMBH presence in NSCs, especially in lower-mass galaxies \citep{Amaro-Seoane:2006,Gurkan:2006,Amaro-Seoane:2007,Sedda:2019c,ASedda:2023}.
However, more massive NSCs (with escape velocities $\gtrsim$400\,km\,s$^{-1}$) inevitably form high-mass IMBHs \citep{Chattopadhyay:2023}.
SMBH formation in spheroids depends on the spheroid's mass, with only those above a certain threshold hosting SMBHs \citep{Kroupa:2020}.
Ultracompact dwarf galaxies may not always possess IMBHs but rather compact sub-clusters of normal black holes and neutron stars \citep{Mahani:2021}.

\subsubsection{Tidal Disruption Events}

Observations of SMBHs are often biased towards detecting the most massive, whereas TDEs only result from $\lesssim10^8\,\mathrm{M_\sun}$ \citep[\textit{cf.}\ $\gtrsim$$10^7\,\mathrm{M_\sun}$, as pointed out by][owing to a population of tidally-destroyed stars that is dominated by low-mass stars]{Coughlin:2022,Nicholl:2022} black holes \citep{Hills:1975,Rees:1988,MacLeod:2012}.
TDEs with IMBHs involve successive close encounters with stars, revealing the IMBH's mass \citep{Fulya:2022}.
TDEs involving white dwarfs and IMBHs \citep[\textit{e.g.},][]{Luminet:1989,Rosswog:2009,Clausen:2011,Haas:2012,Cheng:2014,MacLeod:2016,Vick:2017,Tanikawa:2017,Tanikawa:2018,Maguire:2020,Chen:2022b,Lam:2022} may indicate masses $\lesssim10^5,\mathrm{M_\sun}$ \citep{Kobayashi:2004,Rosswog:2009,Gezari:2021}, aiding in lower-mass black hole identification.
This presents an opportunity to exclusively identify lower-mass black holes and extend black hole mass scaling relations down to lower masses \citep[\textit{e.g.},][]{Ramsden:2022}.
Additionally, gravitational-wave detectors can potentially observe tidal stripping events within the Local Supercluster \citep{Chen:2022b}.

Black hole growth in galactic nuclei can be explained by gas and tidal disruption accretion in dense NSCs \citep{Lee:2022}.
Runaway tidal encounters in NSCs can lead to the formation of SMBHs \citep{Stone:2017,Baldassare:2022}.
TDE rates aid in understanding black hole seeding mechanisms and the BHMF \citep{Stone:2016,Yao:2023,Coughlin:2023}.
Recent observations of TDEs involving IMBHs \citep{Perley:2019,Dacheng:2020,Wen:2021,Angus:2022} contribute to understanding their frequency, with growing samples in dwarf galaxies \citep{Molina:2021}.
TDE rates are higher in nucleated galaxies \citep{Pfister:2020}, but distinguishing between complete and partial TDEs is crucial \citep{Bortolas:2023}.
Infrared surveys complement optical and X-ray surveys for detecting dust-obscured TDEs in star-forming galaxies \citep{Panagiotou:2023}.

\subsubsection{Dwarf Galaxies and Evolution}

Studies examining black holes in dwarf galaxies reveal their crucial role in hierarchical galaxy evolution, indicating a shift from the traditional belief that massive black holes are solely located in giant galaxy nuclei \citep[see][for a review]{Reines:2022}.
Occupancy of IMBHs in dwarf galaxies exceeds 50\% \citep{Greene:2020}, challenging previous notions.
While observational searches for IMBHs in dwarf galaxies are increasing \citep[\textit{e.g.},][]{Dong:2007,Farrell:2009,Reines:2013,Lin:2016,Mateu:2021,Marko:2022,Salehirad:2022,Hatano:2023,Hatano:2023b}, simulations struggle to replicate these findings \citep{Haidar:2022}, highlighting existing discrepancies.
Additionally, feedback from central black holes impacts galaxy evolution, yet inconsistencies persist between simulations \citep{Lanfranchi:2021} and observations \citep{Davis:2022}.

Dwarf galaxies hosting low-luminosity AGNs suggest the presence of accreting IMBHs \citep{Mezcua:2016,Mezcua:2018,Mezcua:2020,Schutte:2022,Yang:2023}.
Understanding black holes in dwarf galaxies is vital for testing formation models and predicting black hole masses \citep{Volonteri:2008,Wassenhove:2010,Zaw:2020,Silk:2017,Barai:2019,Goradzhanov:2022}, especially as evidence suggests dwarf galaxies' black holes may be over-massive compared to larger galaxies \citep[][\textit{cf.}\ \citealt{Pacucci:2018}]{Reines:2011,Secrest:2017,Mezcua:2022,Spinoso:2022,Weller:2023,Stone:2023,Maiolino:2023b,Pacucci:2023,Sanchez:2023}.
The IMBH occupation fraction is crucial for seeding theories \citep{Chadayammuri:2022,Spinoso:2022} and extending black hole mass scaling relations to a lower regime \citep[\textit{e.g.},][]{Graham:2013,Graham:2023a,Graham:2023c,Graham:2023b,Graham:2023d,Savorgnan:2013,Savorgnan:2016,Savorgnan:2016b,Savorgnan:Thesis,Davis:2017,Davis:2018,Davis:2019c,Davis:2019b,Davis:2019,Davis:2021,Davis:2023,Sahu:2019,Sahu:2019b,Sahu:2020,NSahu:2022,NSahu:2022b,Sahu:Thesis,Sahu:2022c,Jin:Davis:2023}.
Improved understanding of black hole--galaxy coevolution aids in calculating the black hole mass density in the Universe \citep[\textit{e.g.},][]{Graham:2007,Davis:2014,Mutlu-Pakdil:2016}, elucidating their evolutionary history, and ascertaining causal mechanisms \citep{Pasquato:2023b}.

\subsection{Future Work and Possible Observations}

In their investigation of the accretion disks around IMBHs, \citet{Ogata:2021} predicted that IMBHs could be observed as bright sources in the infrared band.
Furthermore, \citet{Cann:2018,Cann:2021} have shown that infrared coronal lines can find AGNs even when optical diagnostics fail, as demonstrated in their study of an optically-normal, low-metallicity dwarf galaxy.
As such, coronal lines are ideal and the infrared JWST will be well-suited for spotting IMBHs \citep[][\textit{cf.}\ \citealt{Herenz:2023}]{Reefe:2022,Reefe:2023}.
Armed with superior sensitivity, resolution, and spectroscopic multiplexing capabilities, the ongoing JWST observations will provide a boon to scientific discovery \citep{Kalirai:2018}.

Impressively, the JWST has quickly found confirmed galaxies up to $z=13.2$ \citep{Curtis-Lake:2023} and candidate objects up to an incredible $z\approx20$ \citep{Yan:2023}, as well as surprisingly massive galaxies (up to $\sim$$10^{11}$\,M$_\sun$) at $7.4<z<9.1$ \citep{Labbe:2022}.
\citet{Qin:2023} found that boosted star-formation efficiencies and reduced feedback regulation are necessary to reproduce $z\gtrsim16$ JWST galaxy candidates, which are susceptible to low-redshift contamination from $z\sim5$ galaxies.
See also the cautionary report from \citet{Zavala:2023} that detailed how dusty starbursts at lower redshifts can masquerade as ultra-high photometric redshift galaxies in JWST observations.
Furthermore, it is worth investigating the implications of bursty star formation at high redshifts leading to selection effects and associated biases for the JWST \citep{Sun:2023,Looser:2023b}.
Already testing the onset of spiral galaxy formation, the JWST has helped identify a dusty multiarm spiral galaxy at $z=3.06$ \citep{Wu:2023} and found that disk galaxies are prevalent out to at least $z=9.5$ \citep{Kartaltepe:2022,Sun:2023b}
Similarly, the JWST has enabled the identification of the most distant barred galaxies to-date, out to $z\simeq3$ \citep{Guo:2022,Huang:2023,Costantin:2023}.
Of particular interest to black hole seeding and early BH--galaxy assembly, the JWST has discovered a crop of low-mass \citep[under-massive;][]{Stone:2023b} galaxies at high redshifts \citep{Kocevski:2023,Looser:2023,Gelli:2023,Strait:2023,Curtis-Lake:2023,Robertson:2023,Williams:2023b,Ubler:2023}.

The JWST has the ability to color-select black hole seeds transitioning to SMBHs \citep{Goulding:2022} and distinguish rest-frame optical lines for the identification of ``light seed'' Population~III \citep{Vanzella:2023} and ``heavy seed'' direct-collapse black holes in the early Universe \citep{Nakajima:2022}.
Models estimate that the JWST surveys will have the sensitivity to detect heavy black hole seeds out to redshifts of $z\lesssim14$ \citep{Trinca:2022b}.
Already, the JWST has spotted high-redshift AGNs \citep{Larson:2023,Maiolino:2023,Maiolino:2023b} and established constraints on their formation that requires either super-Eddington accretion from a stellar mass seed or Eddington accretion from a very massive black hole seed.
Moreover, the JWST has already witnessed an X-ray quasar ($M_\bullet\sim4\times10^{7}$\,M$_\sun$ in a comparably massive host galaxy) at $z=10.3$, which ``suggests that early SMBHs originate from heavy seeds'' \citep{Bogdan:2023,Natarajan:2023,Goulding:2023} and constitutes the first outsize (n\'{e}e obese) black hole galaxy \citep{Natarajan:2017}.
Radio observations from the forthcoming Square Kilometre Array could also detect emissions from direct-collapse black holes at high-$z$ \citep{Whalen:2020,Whalen:2023}.
The parameters for black hole growth and seed models will become significantly constrained as future observations discern $z>8$ quasars \citep{Pacucci:2022}.

In order to dynamically confirm the IMBH mass estimates we have made, it would be necessary to resolve kinematics within the gravitational sphere of influence (SOI) of our black holes.
From \citet{Peebles:1972},\footnote{See also the corresponding \S3.3 from \citet{Davis:2020}.} a black hole at the center of a galaxy has an SOI with a radius $r_h\equiv GM_\bullet\sigma_0^{-2}$.
From our sample of 23 targets of interest, UGC~3949 is the only galaxy with a known central velocity dispersion ($\sigma_0=70.1\pm4.9$\,km\,s$^{-1}$).
For UGC~3949 ($d=89.9\pm6.9$\,Mpc and $M_\bullet=5.05\pm2.99\times10^4$\,M$_\sun$), we obtain $r_h = 44.2\pm26.8\,{\rm mpc} = 101\pm62\,\mu{\rm as}$.

With its stunning resolution of $20\,\mu{\rm as}$, the Event Horizon Telescope (EHT) could resolve this region in the center of UGC~3949.
The EHT resolved the emission rings surrounding the SMBHs M87* and Sgr~A$^\ast$ with diameters of $\approx$$42\,\mu{\rm as}$ \citep{EHT,Medeiros:2023} and $\approx$$52\,\mu{\rm as}$ \citep{Akiyama:2022}, respectively.
Indeed, distance alone would not be a hinderance in resolving similarly-sized IMBHs in our sample of targets, which only extends out to about 125\,Mpc.
Granted, even if the SOI were resolved in images of our targets, spectra would still be useful.

We hope to target the 23 candidates with follow-up spectroscopy (for stellar velocity dispersions, gas emission lines, and black hole virial mass estimates\footnote{\citet{Cho:2023} elucidate how better constraints for the H$\alpha$ size--luminosity relation are required to calibrate a virial mass estimator based on the H$\alpha$ broad emission line from low-luminosity AGNs and IMBHs.}), X-ray imaging (for AGNs activity), and perhaps simultaneous radio interferometry \citep[for the fundamental plane of black hole activity;][]{Merloni:2003,Falcke:2004}.
However, \citet{Kayhan:2022} cautioned against the use of the fundamental plane of black hole activity for identifying IMBHs without additional constraints beyond just straightforward X-ray and radio observations.
Furthermore, only a small fraction ($\sim$0.6\%) of IMBH candidate host galaxies are radio-band active \citep{Yang:2023b}.
Further X-ray information such as the characteristic variability or the normalized excess variance (variability amplitude) can also be used to estimate black hole masses of AGNs \citep{Pan:2015}.
However, the characteristic damping timescale at X-ray wavelengths is substantially shorter \citep{Gonzo:2012}, and less correlated with black hole mass, than at optical wavelengths \citep{Burke:2021}.\footnote{Although, \citet{Treiber:2023} revealed the substantial effect of contamination from variable stars in their search for stochastic variability.}
Additionally, the variability timescale at sub-millimeter wavelengths appears to also be a useful parameter that correlates well with black hole mass in low-luminosity AGNs \citep{Chen:2023b}.

\section{Conclusions}\label{sec:conclusions}

The comparisons throughout this work seem to indicate that Sd galaxies exhibit characteristics that are surprisingly similar to the general population of spiral galaxies, in the aggregate, although we did not explore bulge mass, as many Sd galaxies are (considered) bulgeless.
However, we note that a range of $B/T$ flux ratios exist at each morphological type \citep{Graham:2008}.
This unexpected resemblance could be a symptom of subjective morphological misclassifications, or perhaps is showing a natural diversity even amongst this subpopulation of apparent late-type spiral galaxies.
Specifically, we find the expectation that an archetypical Sd should have a low mass, be slowly rotating, display loosely wound spiral arms, and have a low-mass central hole is not unimpeachable; characteristics more akin to earlier type spiral galaxies appear to be endemic.
Thus, the classical Sd morphological class is a stereotype; there is no average Sd galaxy.
In any regard, we do find merit in exploring Sd galaxies for IMBHs, as they possess the requisite environmental traits (for hosting IMBHs) in a higher proportion than the average spiral galaxy.

On the whole, we find that a randomly selected Sd galaxy will have a 27.7\% chance of possessing an IMBH.
Our search has produced 23 candidates, each with a probability of at least 50\% of hosting an IMBH.
Although we expect 23/85 of our galaxies house an IMBH, the product of all their $P(\mathcal{M}_\bullet\leq5)$ probabilities implies an $\approx$100\% certainty that at least one of our 85 Sd galaxies possesses an IMBH.
We intend to make these targets the focus of continued research.
The fruition of the long-sought quest to identify IMBHs will finally complete the gap in our knowledge of the demography of black holes.


\begin{acknowledgments}
BLD thanks David Nelson for the use of his secluded office space during the COVID-19 pandemic.
The Australian Research Council's funding scheme DP17012923 supported this research.
Parts of this research were conducted by the Australian Research Council Centre of Excellence for Gravitational-wave Discovery (OzGrav), through project number CE170100004.
This material is based upon work supported by Tamkeen under the NYU Abu Dhabi Research Institute grant CASS.
This research has made use of NASA's Astrophysics Data System, and the NASA/IPAC Extragalactic Database (NED) and Infrared Science Archive (IRSA).
We acknowledge use of the HyperLeda database (\url{http://leda.univ-lyon1.fr}) and the Reference Catalog of galaxy SEDs (\url{http://rcsed.sai.msu.ru/}).
\end{acknowledgments}

\facilities{
CXO,
GALEX,
Pan-STARRS1, \&
SDSS.
}

\software{
\\
\href{https://github.com/bendavis007/2DFFT}{\textcolor{linkcolor}{\texttt{2DFFT}}} \citep{2dfft}\\
\href{https://github.com/AGES-UARK/2dfft_utils}{\textcolor{linkcolor}{\texttt{2DFFT Utilities}}}\\
\href{https://github.com/ebmonson/2DFFTUtils-Module}{\textcolor{linkcolor}{\texttt{2DFFTUtils Module}}}\\
\href{https://github.com/astropy/astropy}{\textcolor{linkcolor}{\texttt{Astropy}}} \citep{astropy:2013,astropy:2018}\\
\href{https://www.openai.com/chatgpt}{\textcolor{linkcolor}{\texttt{ChatGPT}}}\\
\href{https://github.com/CullanHowlett/HyperFit}{\textcolor{linkcolor}{\texttt{Hyper-Fit}}} \citep{Robotham:2015,Robotham:2016}\\
\href{https://iraf-community.github.io/}{\textcolor{linkcolor}{\texttt{IRAF}}} \citep{Tody:1986,IRAF}\\
\href{http://kcor.sai.msu.ru/}{\textcolor{linkcolor}{\texttt{$K$-corrections calculator}}}\\
\href{https://github.com/matplotlib/matplotlib}{\textcolor{linkcolor}{\texttt{Matplotlib}}} \citep{Hunter:2007}\\
\href{https://github.com/numpy/numpy}{\textcolor{linkcolor}{\texttt{NumPy}}} \citep{harris2020array}\\
\href{https://pandas.pydata.org/}{\textcolor{linkcolor}{\texttt{Pandas}}} \citep{McKinney_2010}\\
\href{https://www.mathworks.com/matlabcentral/fileexchange/26516-pearspdf}{\textcolor{linkcolor}{\texttt{pearspdf}}}\\
\href{https://github.com/MilesCranmer/PySR/tree/v0.12.3}{\textcolor{linkcolor}{\texttt{PySR}}} \citep{Cranmer:2023}\\
\href{https://www.python.org/}{\textcolor{linkcolor}{\texttt{Python}}} \citep{Python}\\
\href{https://sites.google.com/cfa.harvard.edu/saoimageds9}{\textcolor{linkcolor}{\texttt{SAOImageDS9}}} \citep{Joye:2003}\\
\href{https://github.com/scipy/scipy}{\textcolor{linkcolor}{\texttt{SciPy}}} \citep{Virtanen_2020}\\
\href{http://sparcfire.ics.uci.edu/}{\textcolor{linkcolor}{\texttt{SpArcFiRe}}} \citep{sparcfire}\\
\href{https://github.com/DeannaShields/Spirality}{\textcolor{linkcolor}{\texttt{Spirality}}} \citep{spirality}\\
\href{http://pythonhosted.org/uncertainties/}{\textcolor{linkcolor}{\texttt{uncertainties}}}
}

\section*{ORCID iDs}

\begin{CJK*}{UTF8}{gbsn}
\begin{flushleft}
Benjamin L.\ Davis \scalerel*{\includegraphics{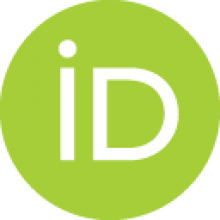}}{B} \url{https://orcid.org/0000-0002-4306-5950}\\
Alister W.\ Graham \scalerel*{\includegraphics{orcid-ID.png}}{B} \url{https://orcid.org/0000-0002-6496-9414}\\
Roberto Soria \scalerel*{\includegraphics{orcid-ID.png}}{B} \url{https://orcid.org/0000-0002-4622-796X}\\
Zehao Jin (金泽灏) \scalerel*{\includegraphics{orcid-ID.png}}{B}\\\url{https://orcid.org/0009-0000-2506-6645}\\
Igor D.\ Karachentsev \scalerel*{\includegraphics{orcid-ID.png}}{B} \url{https://orcid.org/0000-0003-0307-4366}\\
Elena D'Onghia \scalerel*{\includegraphics{orcid-ID.png}}{B} \url{https://orcid.org/0000-0003-2676-8344}
\end{flushleft}
\end{CJK*}

\bibliography{bibliography}

\end{document}